\documentclass{article}
\usepackage[english]{babel}
\usepackage[utf8]{inputenc}
\usepackage{johd}

\usepackage{amsmath,amssymb,amsfonts}
\usepackage{algorithmic}
\usepackage{graphicx}
\usepackage{textcomp}
\usepackage{color,soul}
\usepackage{placeins}
\usepackage{pdflscape}

\newcommand\blfootnote[1]{%
  \begingroup
  \renewcommand\thefootnote{}\footnote{#1}%
  \addtocounter{footnote}{-1}%
  \endgroup
}
\title{Comparison of propagation models and forward calculation methods on cellular, tissue and organ scale atrial electrophysiology\blfootnote{This work has been submitted to the IEEE for possible publication. Copyright may be transferred without notice, after which this version may no longer be accessible.}}
\date{}
\begin{document}
\maketitle
\parbox{\textwidth}{\centering
Claudia Nagel$^{a*}$, Cristian Barrios Espinosa$^{a}$, Karli Gillette$^{b}$, Matthias A.F. Gsell$^{b,c}$, Jorge S\'{a}nchez$^{a}$, Gernot Plank$^{b}$, Olaf Dössel$^{a}$, Axel Loewe$^{a}$ \newline \newline
        \small $^{a}$ Institute of Biomedical Engineering, Karlsruhe Institute of Technology, Kaiserstr. 12, 76131 Karlsruhe, Germany \\
        \small $^{b}$ Division of Biophysics, Medical University of Graz, Neue Stiftingtalstrasse 6/D04, A-8010 Graz, Austria \\
        \small $^{c}$ Institute of Mathematics and Scientific Computing, University of Graz, NAWI Graz, Heinrichstrasse 36, A-8010 Graz, Austria \\
        \small $^{*}$Corresponding author: Claudia Nagel, \tt{publications@ibt.kit.edu}
}

\begin{abstract} 
\noindent \textit{Objective:} The bidomain model and the finite element method are an established standard to mathematically describe cardiac electrophysiology, but are both suboptimal choices for fast and large-scale simulations due to high computational costs. We investigate to what extent simplified approaches for propagation models (monodomain, reaction-eikonal and eikonal) and forward calculation (boundary element and infinite volume conductor) deliver markedly accelerated, yet physiologically accurate simulation results in atrial electrophysiology. 
\textit{Methods:}
We compared action potential durations, local activation times (LATs), and electrocardiograms (ECGs) for sinus rhythm simulations on healthy and fibrotically infiltrated atrial models.
\textit{Results:} 
All simplified model solutions yielded LATs and P~waves in accurate accordance with the bidomain results. Only for the eikonal model with pre-computed action potential templates shifted in time to derive transmembrane voltages, repolarization behavior notably deviated from the bidomain results. ECGs calculated with the boundary element method were characterized by correlation coefficients $>$0.9 compared to the finite element method. The infinite volume conductor method led to lower correlation coefficients caused predominantly by systematic overestimations of P~wave amplitudes in the precordial leads. 
\textit{Conclusion:} 
Our results demonstrate that the eikonal model yields accurate LATs and combined with the boundary element method precise ECGs compared to markedly more expensive full bidomain simulations. However, for an accurate representation of atrial repolarization dynamics, diffusion terms must be accounted for in simplified models. 
\textit{Significance:} Simulations of atrial LATs and ECGs can be notably accelerated to clinically feasible time frames at high accuracy by resorting to the eikonal and boundary element methods. \end{abstract}

\noindent\\

\section{Introduction}
\label{sec:introduction}
In computational cardiac modeling, the bidomain model is the most biophysically detailed formulation to compute the spread of the de- and repolarization wavefront and the electrical source distribution throughout the cardiac tissue. Furthermore, the finite element method is considered the gold standard for computing the body surface potentials from a given distribution of the electrical sources in the heart to extract electrocardiograms (ECG) at standardized electrode positions. However, both methods are computationally expensive and are thus suboptimal for generating large \textit{in silico} datasets of cardiac signals as they are required e.g. for machine learning applications~\cite{Nagel-2021-ID16114, Luongo-2022-ID17367}, or for simulating excitation propagation close to real time on cardiac digital twins to e.g. guide ablation therapy~\cite{Azzolin-2021-ID16137}. Hence, simplified models with fast solution times are needed to speed up the generation of \textit{in silico} datasets of cardiac signals, such as local activation times (LATs), electrograms or ECGs by several orders of magnitude~\cite{wallman12,neic17}. Yet, the signals obtained with these simplified methods need to be physiologically accurate and resemble the results obtained with the bidomain and finite element method. In this work, we therefore aim to quantify the inaccuracies arising in simulated atrial signals when resorting to simplified computational methods. While comparisons of this type have already been performed for the ventricles~\cite{neic17, Bishop-2011-ID12010} and partly also for four chamber heart models~\cite{Potse-2006-ID15641}, a study focusing on atrial electrophysiology is lacking to the best of our knowledge. However, this is substantial since the atria stand out by a highly complex myocardial fiber structure, locally heterogeneous properties regarding ion channel and tissue conductivities and higher anisotropy ratios as compared to the ventricles.  

The monodomain, reaction-eikonal, and the eikonal model solved by the fast iterative method constitute the simplified propagation models investigated in this work. Forward calculation techniques applied in this study comprise the boundary element and the infinite volume conductor methods. Simulations were carried out in sinus rhythm with and without the inclusion of fibrotic tissue modeled as passive conduction barriers~\cite{vigmond16}, slow conducting tissue patches and rescaled ion channel conductivities representing cytokine effects~\cite{Azzolin-2021-ID16653, Roney-2016-ID12208}. We assess the errors between simplified propagation models and forward calculation methods to the gold standard bidomain and finite element formulations with metrics extracted from the simulation results on cellular, tissue and organ scale comprising APDs, LATs, and ECGs, respectively. 

\section{Methods}
\subsection{Model Generation}
\label{sec:anatomicalModel}
\begin{figure*}[!h]
\centerline{\includegraphics[width = \textwidth]{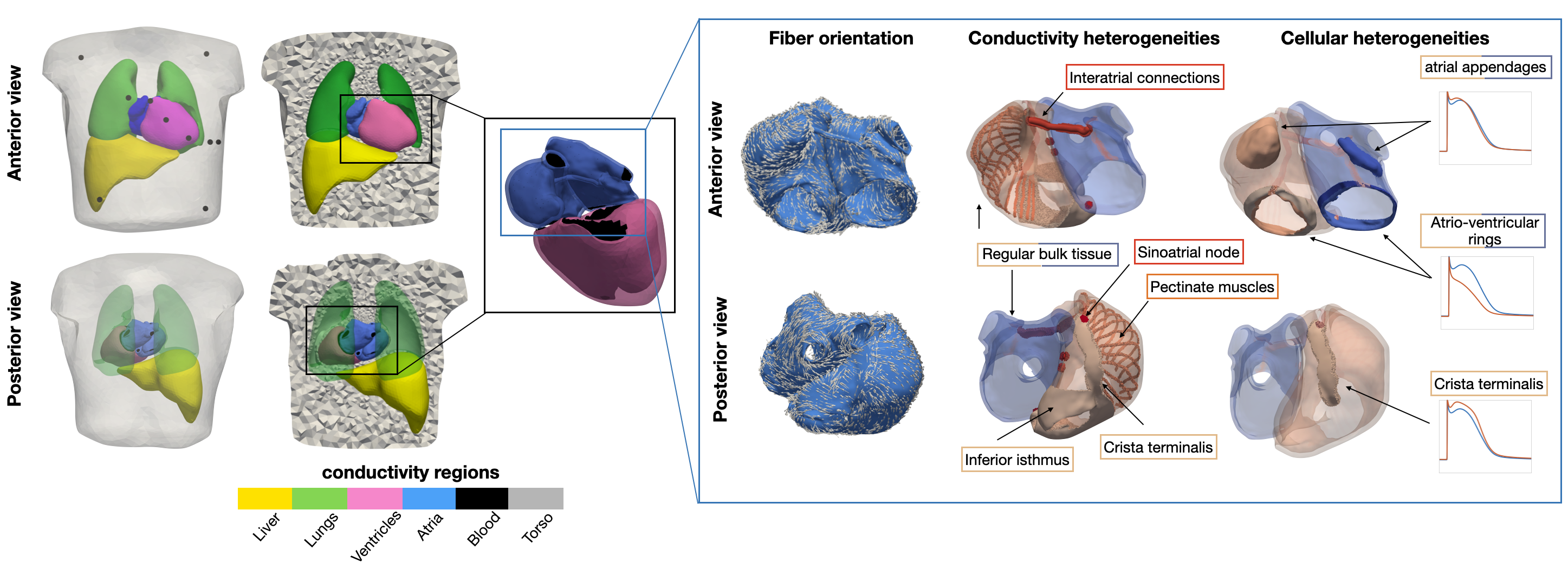}}
\caption{Torso model with the segmented organs and electrode positions (transparent and clipped views) from the anterior and posterior view in the top and bottom row, respectively. The right panel shows the anatomically detailed atrial model that was augmented with fiber orientation and labels for anatomical structures. Heterogeneous conductivity and ionic properties were assigned to spatially distinct regions in the mesh. Resulting APs featuring ionic heterogeneity are depicted on the right side in red together with the baseline Courtemanche \textit{et al.} cellular model in blue.}
\label{anatomicalModel}
\end{figure*}

An anatomically detailed model of the torso was obtained by multi-label magnetic resonance image segmentation as described in~\cite{krueger13b}. The contours of atria, ventricles, lungs, liver and torso were exported as triangular surface meshes. These were smoothed and resampled with an average edge length of 0.5\,mm, 5\,mm, 5\,mm, 7\,mm and 15\,mm, respectively, using \textit{Meshmixer} (Autodesk, San Rafael, CA, USA) and \textit{InstantMeshes}~\cite{Jakob-2015-ID12648} whereby details were corrected manually in \textit{Blender} (Blender Foundation, Amsterdam, The Netherlands) to avoid intersecting segments and ensure a sufficient mesh quality and topology. The segmented atrial endocardial surfaces were fed into the pipeline described in~\cite{Azzolin-2021-ID16653,Zheng-2021-ID17236} to obtain a volumetric tetrahedral bi-atrial geometry with a homogeneous wall thickness of 3\,mm and an average edge length of 523\,$\mu$m augmented with inter-atrial connections, labels for anatomical structures and myocardial fiber orientation. \textit{Meshtool}~\cite{Neic-2020-ID13697} was used to generate a tetrahedral model of the full torso while preserving the surfaces of the considered organs. Tags for the atrial and ventricular blood pools were allocated to all elements in the volumetric torso model located inside the surfaces bounded by the atrial and ventricular endocardial walls with closed valve and vein orifices. A detailed view of the torso and atrial model is depicted in Fig.~\ref{anatomicalModel}. 

Isotropic extracellular conductivity of 0.0389\,S/m, 0.028\,S/m, 0.06\,S/m, 0.7\,S/m and 0.22\,S/m was assigned to lungs, liver, ventricles, atrial and ventricular blood pools and the remaining torso tissue, respectively, as reported in previous work~\cite{Gillette-2021-ID16142, reinke14}. In order to conduct comparable experiments with the mono- or bidomain model that require conductivities and the (reaction-)eikonal models that resort to conduction velocities (CVs) being assigned to spatially distinct areas in the atria, it is crucial to consistently associate conductivities and CVs for all heterogeneous tissue regions. Therefore, anisotropic and locally heterogeneous conductivities were assigned to five different regions in the atria comprising regular bulk tissue, crista terminalis, pectinate muscles, inferior isthmus, and inter-atrial connections as follows: 
CVs corresponding to the monodomain conductivities reported in~\cite{loewe15b} for 0.33\,mm resolution voxel models were calculated as described in~\cite{krueger13e}. Using \textit{tuneCV}~\cite{Plank-2021-ID15953, costa-cinc-2010}, intra- ($\sigma_i$) and extracellular ($\sigma_e$) conductivities as well as the monodomain conductivities ($\sigma_m$) were iteratively optimized for the tetrahedral mesh setup described above while keeping the $\sigma_i$/$\sigma_e$ ratio fixed. For this purpose, five strand geometries with a length of 10\,cm were generated each characterized by a resolution corresponding to the average edge length of one of the heterogeneous conductivity regions in the atria. Intra- and extracellular conductivities in longitudinal and transversal fiber direction as reported by Clerc \textit{et al.}~\cite{Clerc-1976-ID12471} as well as by Roberts \textit{et al.}~\cite{roberts82} were initially assigned to the elements in the slab meshes. In an iterative optimization procedure, the conductivities were adjusted until the CV converged to the target value derived from~\cite{loewe15b}. In this way, the originally reported intra- and extracellular conductivity values were scaled while the ratios between them were kept constant along the eigenaxes~\cite{costa-cinc-2010}. In the following, we refer to the tuned conductivities obtained by initially assigning the values reported by Clerc~\cite{Clerc-1976-ID12471} and Roberts \textit{et al.}~\cite{roberts82} to the slab meshes as Clerc and Roberts conductivities, respectively. The resulting heterogeneous and anisotropic conductivity setup for each atrial region is summarized in table SI in the supplementary material. For the monodomain simulations, we considered two different cases which we refer to as "monodomain with and without explicit conductivity tuning". For the first one, we repeated the \textit{tuneCV} optimization using the monodomain propagation model and obtained the monodomain conductivities listed in table SI in the supplementary material. In the second case, we directly computed the monodomain conductivities from the tuned intra- and extracellular bidomain conductivities as half of their harmonic mean. 

The Courtemanche \textit{et al.} cell model~\cite{Courtemanche-1998-ID17067} was used in the simulations described in section~\ref{sec:propagationDrivers}. To reflect regionally heterogeneous electrophysiology, maximum ion channel conductances were rescaled compared to the baseline model as reported in previous work~\cite{loewe15b, krueger13} and are summarized in table SII in the supplementary material. The final CV values in longitudinal and transversal fiber direction as used for the eikonal and reaction-eikonal simulations described in section~\ref{sec:propagationDrivers} were subsequently calculated with \textit{tuneCV}~\cite{Plank-2021-ID15953} based on the tissue and ion channel conductivity settings in each atrial region.

\subsection{Propagation Models}
\begin{figure*}[!h]
\centerline{\includegraphics[width = \textwidth]{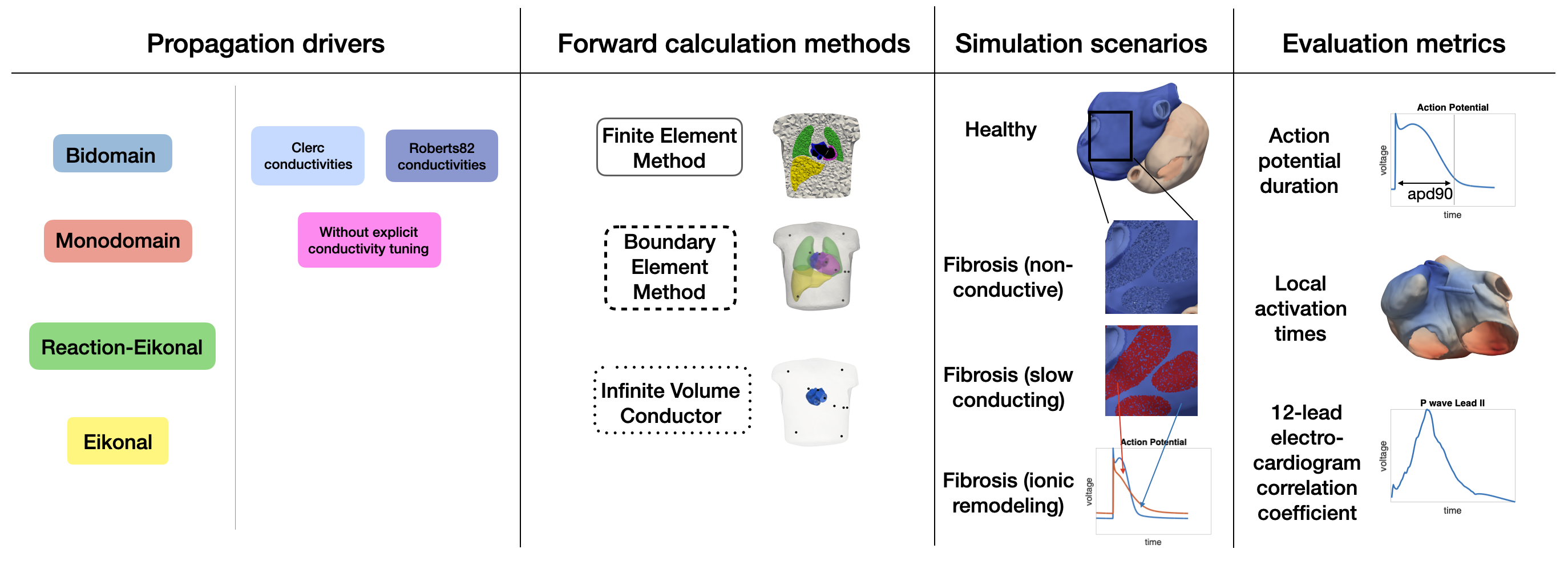}}
\caption{Overview of the different propagation models, forward calculation methods, simulation scenarios and evaluation metrics used in this work.}
\label{overview_methods}
\end{figure*}

\label{sec:propagationDrivers}
\subsubsection{Bidomain Model}

The bidomain model represents the propagation of the electrical de- and repolarization wavefront through the cardiac tissue~\cite{tung1978bi,franzone2014mathematical,keener2009mathematical}. Here, the intracellular and extracelullar domains are coupled and intertwined in a system of partial differential equations. Solving this system provides at each point in the cardiac tissue the intracellular $\Phi_i$ and extracellular $\Phi_e$ electrical potentials:
\begin{equation} \label{eq1}
-\nabla \cdot (\sigma_i + \sigma_e) \nabla \Phi_e = \nabla \cdot \sigma_i \nabla V_m 
\end{equation}
\begin{equation} \label{bidomain_eq1}
\nabla \cdot (\sigma_i\nabla V_m) + \nabla \cdot (\sigma_i \nabla \Phi_e) = \beta(C_m \frac{\partial V_m}{\partial t} + I_{ion}-I_s) 
\end{equation}
where $V_m=\phi_i-\phi_e$ is the transmembrane potential  $\sigma_i$ and $\sigma_e$ are the intracellular and extracellular conductivity tensors, respectively,  $\beta$ is the surface to volume ratio of the membrane, and $C_m$ is the membrane capacitance per unit surface. Additionally, $I_{s}$ and $I_{ion}$ are the transmembrane stimulus and ionic currents, respectively. The ionic current $I_{ion}$ depends on the state variables $\eta$ that are calculated with a non-linear system of ordinary differential equations. At the intracellular domain, certain boundary conditions are enforced: $\phi_e$ is constrained to be continuous and there is non-flux in the normal direction for $\phi_i$. At the boundary of the torso and the boundary of the bath that is not in contact with the intracelullar domain,  a non-flux boundary condition for $\phi_e$ is also imposed. Numerical methods used to solve the bidomain model equations rely on high resolution meshes which is the main cause of the model's high computational cost~\cite{vigmond2008solvers}. Nonetheless, the bidomain model is considered the most accurate of the available cardiac models for tissue level electrophysiology.

\subsubsection{Monodomain Model}
Assuming that the intra- and extracellular conductivity tensors are proportional, i.e. their anisotropic ratios are equal, the bidomain model can be significantly reduced to the monodomain model~\cite{keener2009mathematical,franzone2014mathematical,vigmond2008solvers}:

\begin{equation} \label{eq3}
\begin{split}
\beta C_m \frac{\partial V_m}{\partial t} = \nabla \cdot \sigma_m\nabla V_m -\beta (I_{ion}(V_m,\eta)-I_s)
\end{split}
\end{equation}
where $\sigma_m$ is the monodomain conductivity tensor. This tensor can be expressed in terms of half the harmonic mean of intra- and extracellular conductivity tensors: 
\begin{equation} \label{eq2}
\begin{split}
\sigma_m=\frac{\sigma_i\sigma_e}{\sigma_i +\sigma_e}
\end{split}
\end{equation}
The assumption of equal anisotropy does not fully hold in reality. However, this model still offers a close approximation of the wave propagation.~\cite{Potse-2006-ID15641}. For a planar wave moving along the fiber directions monodomain and bidomain models are exactly equivalent. The extracellular potential field $\Phi_e$ can be approximated from the monodomain transmembrane potentials as a source model by solving the elliptic bidomain equation~(\ref{eq1}) at a temporally coarser scale. However, the volume conductor cannot influence the transmembrane voltage distribution in this approach and bath loading effects are ignored. This concept is referred to as pseudo-bidomain approach~\cite{bishop2011bidomain} and is computationally only marginally more expensive than a standard monodomain simulation. 
The monodomain and bidomain models can be discretized in space using different approaches~\cite{niederer2011verification}. For this study, we used the finite element method~\cite{Plank-2021-ID15953}.

\subsubsection{Eikonal Model}
The eikonal model is based on the macroscopic kinetics of the wavefront propagation~\cite{keener2009mathematical,neic2017efficient,pernod2011multi,franzone2014mathematical}. Solving the eikonal equation seeks to find the activation times $T$ for each node based on a local speed function: 

\begin{equation} \label{eq4}
\begin{split}
\sqrt{\nabla T^{\top} M \nabla T} = 1~,
\end{split}
\end{equation}
where $M$ is the squared CV tensor. Although $V_m$ is not directly calculated in this model, it can be inferred from the activation times:
\begin{equation} \label{eq4_1}
\begin{split}
V_m(x,t)=U(x,t-T(x))
\end{split}
\end{equation}

where U is a space-varying AP shape.  Numerical simulations are significantly faster because of the simplicity of the equation and lower resolution meshes that are required. Unfortunately, the eikonal model fails to accurately represent the influence of bath loading effects, high wavefront curvatures, reentry, and wave-collisions on CV.~\cite{neic2017efficient,pullan2002finite}. In the case of complex patterns of activation that occur for example during atrial fibrillation, these limitations become more significant. Nonetheless, these simulations still provide a decent approximation of wave propagation under healthy conditions. In these case, the shortcomings of the eikonal model are still present but their effects are less pronounced.

\subsubsection{Reaction-Eikonal Model}
When applied to coarse meshes, the mono- and bidomain models both exhibit slowed CV for a given tissue conductivity~\cite{niederer2011verification,neic2017efficient}. The reaction-eikonal model uses the activation times obtained by the eikonal model to enable biophysical models to calculate the transmembrane potential in coarse meshes~\cite{neic17}. The resolution requirement is relaxed because the thin wavefront does not need to be explicitly represented. The RE model calculates an $I_{foot}$ current to replicate the effect of the diffusion term and applies it to the reaction model at the time given by the eikonal model. In this work, only the RE$^+$ version of the model is considered. In this version, the $I_{foot}$ current is added to the diffusion term instead of replacing it. Thus, the activation of the nodes can be achieved by either the diffusion term or the $I_{foot}$ current and neighboring nodes also interact during repolarization. The RE$^+$ variant is more accurate when comparing to the monodomain model in coarse meshes and the repolarization gradients are significantly smoother. However, RE models are unable to activate the same node several times (as for example required for simulations of reentry) and share the limitations of the standard eikonal model regarding the influences of wavefront curvature, source sink mismatch and bath loading on CV. 

\subsection{Forward Calculation Methods}
When modeling the torso as a passive volume conductor, the bidomain formulation can be reduced to its parabolic part to solve the forward problem of electrocardiography for a given distribution of $V_m$. The Poisson equation in equation (\ref{eq1}) can be solved numerically by discretizing the full torso domain with finite elements (finite element method). Standard extracellular conductivities were hereby assigned to different organs as described in section~\ref{sec:anatomicalModel}. By assuming isotropic myocardial properties in the extracellular space, a reduced set of dipole sources can be mapped onto the surfaces bounding the organs with different conductivity properties. Then, the boundary element method can be used for computing the body surface potentials and the ECG. In the latter case, applying Green formulas and boundary conditions as well as assuming equal anisotropy ratios in the intra- and extracellular domain allow for reformulating equation (\ref{eq1}) as a surface integral to compute the extracellular potential $\Phi$ at any point $\Vec{r}$ on the torso surface:
\begin{equation}
    \Phi(\Vec{r}) \approx \frac{2(\sigma_i+\sigma_e)}{\sigma_T} \Phi^{\infty}(\Vec{r}) \\- \frac{1}{2\pi} \sum_{k = 1}^K \frac{\sigma_-^k - \sigma_+^k}{\sigma_T} \int_{S^k} \Phi(\Vec{r}) \nabla\left(\frac{1}{r}\right) \Vec{dS'},
    \label{eq:bem}
\end{equation}
whereby the minuend describes the potentials attributable to the sources in an unbounded medium with conductivity  $\sigma_T$. The subtrahend in (\ref{eq:bem}) accounts for secondary sources introduced by the bounded volume conductor. $\sigma_-^k$ and $\sigma_+^k$ characterize the conductivities inside and outside the respective surface $k$. The potential $\Phi^\infty$ can be expressed either using the transmembrane voltage distribution on the cardiac surface~\cite{geselowitz83} or the primary impressed currents~\cite{stenroos07} as volumetric sources:

\begin{equation}
    \Phi^\infty(\Vec{r}) = -\frac{1}{4\pi(\sigma_i+\sigma_e)} \int_{V'} \sigma_i \nabla V_m \left(\frac{1}{r}\right) dV'
\end{equation}

When computing the ECG using the infinite volume conductor method (IVC), the heart is assumed to be immersed in a medium of infinite spatial extent with a homogeneous conductivity $\sigma_T$. This reflects in the sum over the surface integrals in (\ref{eq:bem}) being neglected for calculating extracellular potentials on the torso surface:
\begin{equation}
    \Phi(\Vec{r}) \approx \frac{1}{2\pi \sigma_T} \Phi^{\infty}(\Vec{r})
\end{equation}

\subsection{Simulation Scenarios}
Simulations were carried out on the bi-atrial volumetric model described in section~\ref{sec:anatomicalModel} in sinus rhythm with and without the inclusion of fibrotic tissue patches. For the former case, several elliptically shaped patches with their principal axis aligned to the macroscopic atrial fiber orientation were manually defined predominantly on the posterior wall of the left atrium and the left pulmonary vein antrum as reported by Highuchi \textit{et al.}~\cite{Higuchi-2018-ID11688}. These regions extended transmurally and are shown in Fig.~\ref{overview_methods}. To not only account for the patchiness of atrial fibrosis but also for its diffuse deposition, 80\,\% of the cells within the elliptical patches were defined as fibrotic. In this way, the volume fraction of left atrial fibrosis quantified to 22\,\% of the total left atrial tissue volume. 
Remodeled conduction properties were assigned the fibrotic regions in three different ways: In the first case, fibrotic elements were removed from the atrial mesh and instead assigned to the extracellular domain following the concept of percolation~\cite{vigmond16}. In this way, we introduced passive conduction barriers that do not have a transmembrane voltage and thus do not contribute to the electrical source distribution on the myocardial tissue surface. In the second case, fibrotic regions were characterized as slow conducting patches with CVs reduced by 80\,\% in transversal and 50\,\% in longitudinal fiber direction compared to the healthy baseline case. Conductivities in these regions were obtained as described in section~\ref{anatomicalModel}. In this way, anisotropy ratios were increased by a factor of 2.5 in fibrotic areas promoting wave propagation along myocardial fiber orientation and thus forming the basis for functional reentry circuits. In the third case, ionic properties of the fibrotic cells were remodeled by rescaling the conductances of the sodium ($g_{Na}$), the L-type calcium ($g_{CaL}$) and the inward rectifier potassium current ($g_{K1}$) by a factor of 0.6, 0.5 and 0.5, respectively, compared to the baseline conductances of the Courtemanche \textit{et al.} cell model to account for cytokine remodeling effects~\cite{Roney-2016-ID12208}.

Sinus rhythm simulations were initiated at a sinus node exit site located at the junction of crista terminalis and the superior vena cava. To obtain the transmembrane voltage distribution for the LATs computed with the eikonal model, we used precomputed AP templates extracted from simulations on the same strand geometries as described in section~\ref{anatomicalModel}, whereby the respective ionic model parameters in each region as listed in table SII in the supplementary material were taken into account. These APs were shifted in time according to the computed LAT map. 

The Cardiac Arrhythmia Research Package (\textit{CARP})~\cite{Vigmond-2003-ID14313} and \textit{openCARP}~\cite{Plank-2021-ID15953} were used for computing the spread of the depolarization wave with different propagation models as well as ECGs with the finite element and the infinite volume conductor method. The algorithms described by Stenroos \textit{et al.}~\cite{stenroos07} were used for calculating ECGs with the boundary element method. As recommended by Schuler \textit{et al.}~\cite{Schuler-2019-ID12702}, we downsampled the surface mesh bounding the atria to a resolution of 2.5\,mm for computing the transfer matrix. Furthermore, we applied Laplacian smoothing to the transmembrane voltage sources to ensure a continuous wave propagation on the coarse mesh.

\subsection{Evaluation Metrics}
From the source distribution obtained from simulations using different propagation models, we calculated APDs at 90\,\% repolarization (APD$_{90}$) for each node in the mesh. Also at each vertex in the geometry, we extracted LATs defined as the timestep with the steepest AP upstroke. For both, APDs and LATs, the accuracy of each propagation model simulation was quantified as the absolute difference to the respective value for each metric obtained from the bidomain simulation with the Clerc conductivities.

To assess fidelity of simplified forward calculation methods along with different propagation models, we evaluated the Pearson correlation coefficient of the respective ECG results with the ECGs obtained by solving the forward problem with the finite element method based on the bidomain source model computed with the Clerc conductivities.

\section{Results}
\subsection{Propagation Models}
\begin{figure*}[!h]
\centerline{\includegraphics[width = \textwidth]{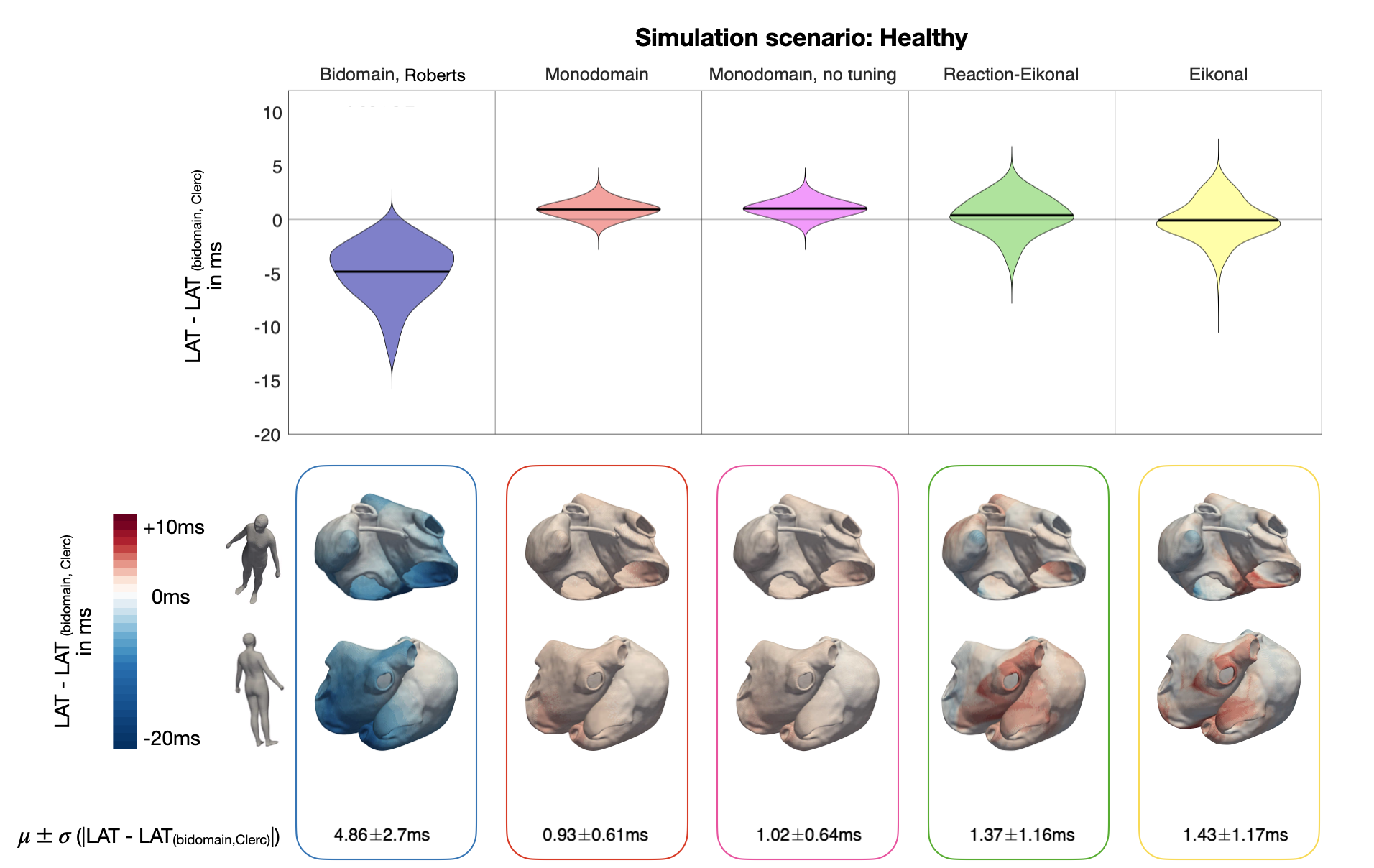}}
\caption{Local activation time (LAT) results in sinus rhythm for healthy (non-fibrotic) tissue for different propagation models. The top panel shows the distribution of the signed LAT differences taking the bidomain simulations executed with the Clerc conductivity ratios as a reference. From left to right, the violin plots show the results for the bidomain (Roberts conductivities), the monodomain (with and without explicit conductivity tuning), the reaction-eikonal and the eikonal simulations. The bottom panel shows the signed LAT differences mapped on the atrial geometry for each propagation model in the above order. Mean and standard deviation of the absolute LAT differences are shown in the bottom row for each case.}
\label{results_lats}
\end{figure*}

The effect of different propagation models on the activation sequence (LATs) is visualized in Fig.~\ref{results_lats}. In the top panel, the distributions of the signed differences between the examined propagation models' LATs and the bidomain results obtained with Clerc conductivity ratios evaluated at all mesh nodes are visualized as violin plots. In the bottom panel, the difference to the bidomain results are mapped onto the atrial geometry. The maximum activation time was 105\,ms for the bidomain simulation with the Clerc conductivities.  
The mismatch in LATs was most pronounced for the bidomain scenario with Roberts conductivities and much smaller for the simplified propagation models. For the Roberts conductivity ratios, the LATs were systematically smaller than the ones resulting from the reference bidomain simulation with the Clerc conductivity settings. Furthermore, the error increased with the spread of the depolarization wave front leading to small deviations close to the sinus node exit site, but errors of up to -14\,ms at the latest activated areas at the posterior wall of the left atrium and the coronary sinus in the right atrium.
The mean and standard deviation of the absolute errors between the bidomain and monodomain LATs with and without explicit conductivitiy tuning were 0.93$\pm$0.61\,ms and 1.02$\pm$0.64\,ms. With the temporal resolution of the sampled simulated myocyte APs being 1\,ms and the LATs being calculated as the point in time marking the steepest AP upstroke, in particular the LAT results for the monodomain simulation with additional conductivity tuning were below the accuracy with which the LATs were determined. 
Reaction-eikonal and eikonal LAT differences quantified to 1.37$\pm$1.16\,ms and 1.43$\pm$1.17\,ms, respectively. The signed LAT error to the bidomain results was distributed similarly across the atrial tissue among these two propagation models (see Fig.~\ref{results_lats} bottom panel).
The LAT results in the simulation scenarios involving fibrosis remodeling were only slightly different compared to the sinus rhythm results depicted in Fig.~\ref{results_lats}. The largest differences occurred for the eikonal propagation model in the simulation scenario where fibrosis was modeled as slow conducting tissue. There, the absolute error to the bidomain results quantified to 1.71$\pm$1.46\,ms compared to 1.43$\pm$1.17\,ms in sinus rhythm without the inclusion of fibrosis. 

\begin{figure*}[!h]
\centerline{\includegraphics[width = \textwidth]{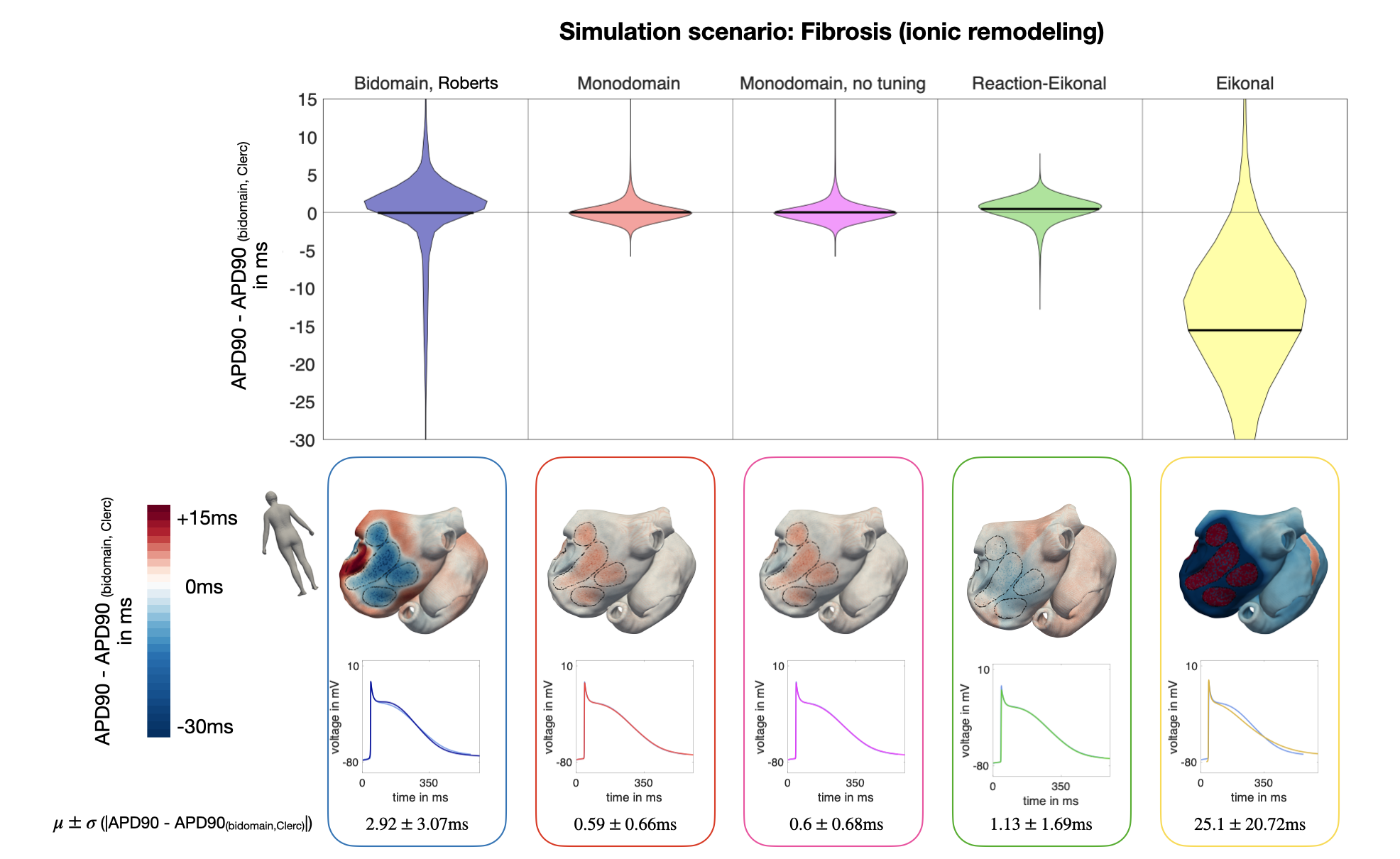}}
\caption{APD$_{90}$ results in sinus rhythm with fibrotic substrate replacing 22\% of the left atrial myocardial tissue modeled as rescaled ionic conductances. The violin plots in the top panel represent the distribution of APD$_{90}$ discrepancies to the bidomain results for all investigated propagation models. In the bottom panel, the signed APD$_{90}$ differences are mapped onto the atrial geometry. Fibrotic regions are encircled with black dashed lines. The APs are shown for one node within the fibrotic area on the posterior left atrial wall. Bidomain APs are visualized in light blue, the other signal trace was obtained with the respective propagation model. The numbers in the bottom line show the mean and standard deviation of absolute APD$_{90}$ differences with respect to the bidomain simulation results.}
\label{results_apd90}
\end{figure*}

APD$_{90}$ results are visualized for the simulation scenario with fibrosis modeled as ionic conductance rescaling in Fig.~\ref{results_apd90}. For the monodomain simulations, the mean and standard deviation of the absolute APD$_{90}$ discrepancies to the bidomain results obtained with Clerc conductivity ratios were below the temporal resolution of the AP time course of 1\,ms. Absolute errors to the bidomain simulation with Roberts conductivity ratios and the reaction-eikonal results quantified to 2.92$\pm$3.07\,ms and 1.35$\pm$1.69\,ms, respectively. In both cases, the highest errors occurred in regions around the fibrotic tissue patches. APD$_{90}$ results for the eikonal simulation were characterized by an absolute error to the bidomain simulation results of 25.1$\pm$20.72\,ms. Furthermore, the AP signal trace obtained from a tissue strand simulation and used as a template to infer the transmembrane voltage distribution for the eikonal LATs is visually clearly distinguishable from the bidomain AP especially in fibrotic regions (see Fig.~\ref{results_apd90} bottom panel).  

\begin{figure*}[!h]
\centerline{\includegraphics[width = \textwidth]{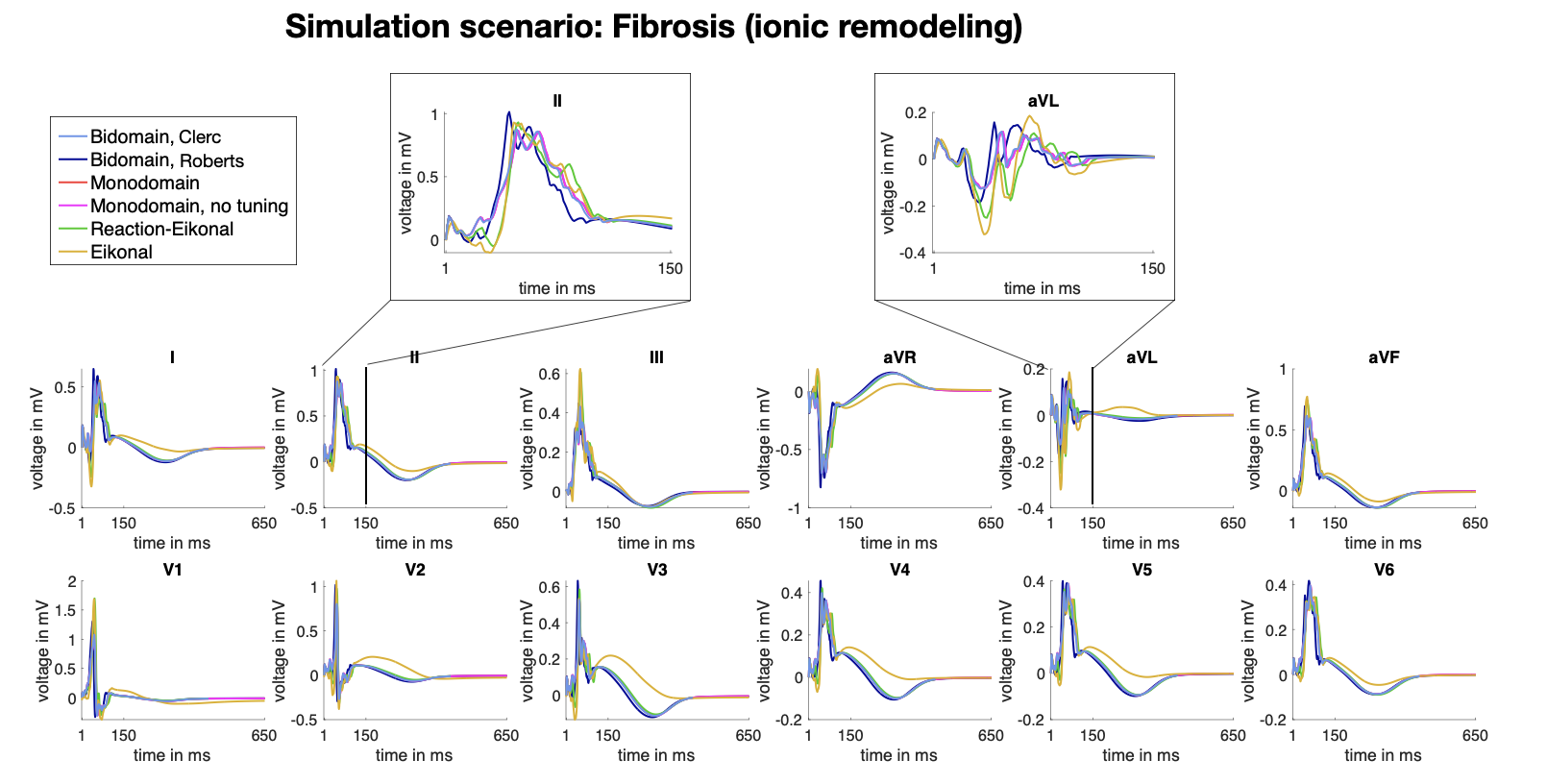}}
\caption{ECGs calculated with the same forward calculation method (BEM) but different propagation models (color coded) with the transmembrane voltages resulting from the simulation scenario with fibrosis modeled as ionic conductivity rescaling.}
\label{results_ecg_differentPropagationDrivers}
\end{figure*}

ECGs obtained from the transmembrane voltage distributions from the simulation scenario with fibrosis modeled as ionic rescaling as depicted in Fig.~\ref{results_apd90} and using the boundary element forward calculation method are visualized in Fig.~\ref{results_ecg_differentPropagationDrivers}. The 12-lead ECG is displayed for a duration of 650\,ms whereby the signal sections in the interval [0\,ms, 150\,ms] and [150\,ms, 650\,ms] represent the P~wave and the atrial repolarization, respectively. The latter is typically not visible in the ECG of a full heartbeat since the repolarization phase of the atria temporally coincides with the ventricular activation and the respective signal parts are thus buried within the QRS~complex. 

The observed discrepancies in the AP signal course between the bidomain and eikonal simulation also reflects in the ECG. As can be seen in Fig.~\ref{results_ecg_differentPropagationDrivers}, the repolarization signal obtained with the eikonal and bidomain propagation model differ. In lead aVL, the polarity of the repolarization wave was even inverted. Apart from the atrial repolarization ECG signal obtained with the eikonal model and precomputed AP templates, the choice of the propagation model did not markedly influence the ECG as the remaining signals in Fig.~\ref{results_ecg_differentPropagationDrivers} show only minor differences. Furthermore, the correlation coefficients between the bidomain ECG obtained with the Clerc conductivity ratios and the other examined propagation models are summarized in Table~\ref{tab:corCoeff_diffPG_sameFWC} for the intervals [0\,ms, 150\,ms] (P~wave), [150\,ms, 650\,ms] (repolarization) and [0\,ms, 650\,ms]. The lowest correlation coefficient for the P~wave occurred for the bidomain simulation with Roberts conductivity ratios. For all simplified propagation models, the P~wave correlation coefficients were above 0.92. Except for the eikonal model, the correlation coefficient of the ECG signal sections representing the repolarization phase wave was above 0.99. 
\begin{table}[]
\caption{Correlation coefficients between the ECGs obtained with the bidomain and simplified propagation models when solving the forward problem with BEM. Columns represent depolarization (P~wave), repolarization and the entire signal. }
\begin{scriptsize}
\begin{center}
\begin{tabular}{l|l|l|l}
                      & {[}0\,ms, 150\,ms{]} & {[}150\,ms, 650\,ms{]} & {[}0\,ms, 650\,ms{]} \\ \hline
Bidomain, Clerc       & 1                    & 1                      & 1                    \\ \hline
Bidomain, Roberts   & 0.8590               & 0.9904                 & 0.9028               \\ \hline
Monodomain            & 0.9942               & 0.9998                 & 0.9961               \\ \hline
Monodomain, no tuning & 0.9931               & 0.9998                 & 0.9954               \\ \hline
Reaction-Eikonal      & 0.9211               & 0.9953                 & 0.9428               \\ \hline
Eikonal               & 0.9203               & 0.6233                 & 0.8791              
\end{tabular}

\label{tab:corCoeff_diffPG_sameFWC}
\end{center}
\end{scriptsize}
\end{table}

ECG and APD$_{90}$ results only marginally differed for the remaining fibrosis remodeling scenarios as detailed and visualized in the figures S6-S10 in the supplementary material.

\subsection{Forward Calculation Methods}
\begin{figure*}[!h]
\centerline{\includegraphics[width = \textwidth]{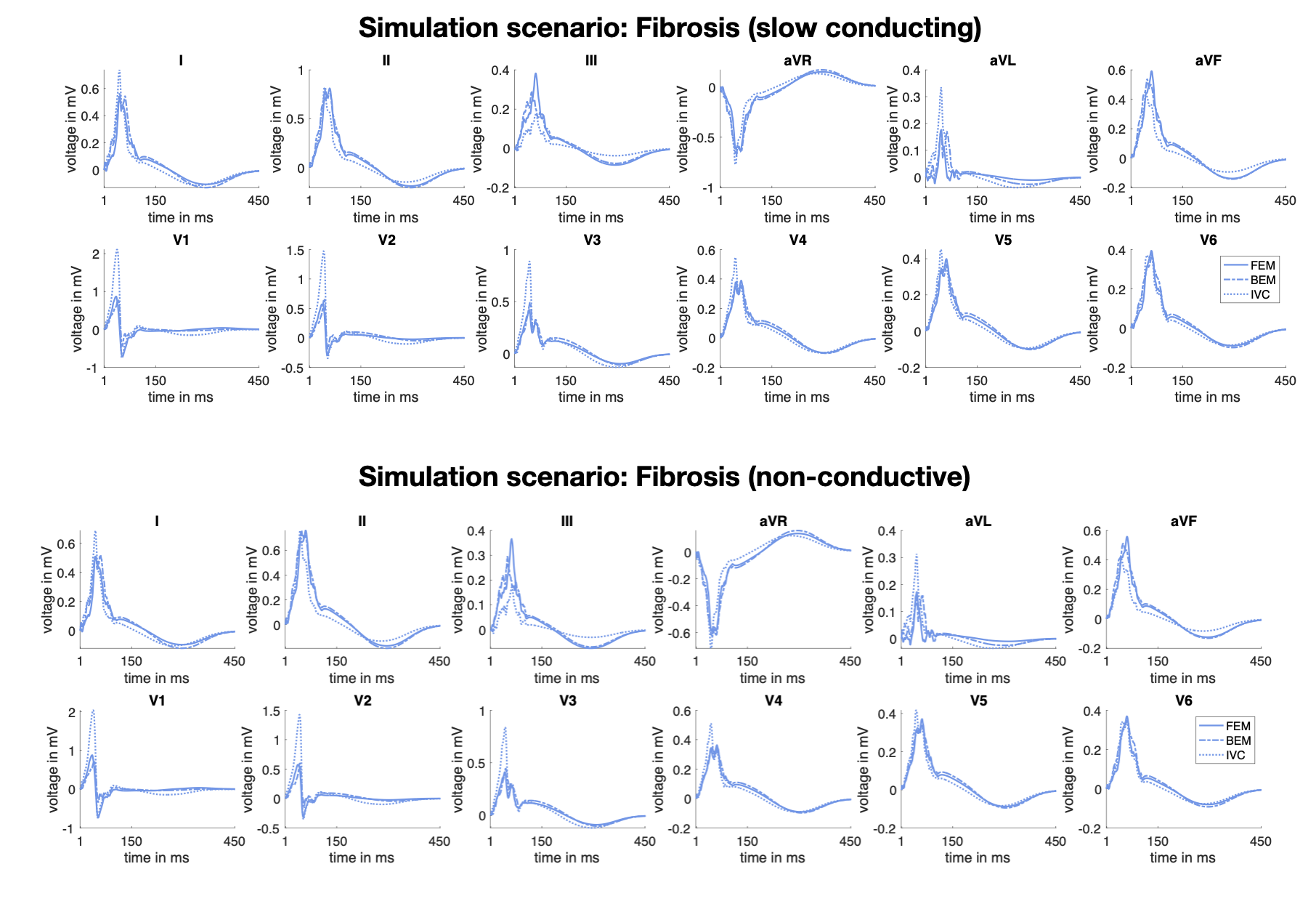}}
\caption{ECGs calculated the same source distribution resulting from a bidomain simulation with Clerc conductivity ratios and with different forward calculation methods in the simulation scenario with fibrosis modeled as slow conducting tissue (top panel) and non-conductive propagation barriers (bottom panel). ECGs calculated with the finite element method (FEM), the boundary element method (BEM) and the infinite volume conductor method (IVC) are visualized with the solid, dash-dotted and dotted lines, respectively.}
\label{ecg_diffFWC}
\end{figure*}

\color{black}
ECGs calculated with different forward calculation methods based on the same source distribution stemming from the bidomain simulation with Clerc conductivity ratios are depicted in Fig.~\ref{ecg_diffFWC} for the simulation scenarios with fibrosis modeled as slow conducting and non-conductive tissue. 
Subtle changes, such as a decrease in V6 amplitudes for example, characterize the differences in the ECGs in the top and the bottom panel. These characteristics are discussed in detail in section~\ref{sec:discussion}.
The correlation coefficients covering the ECG signal parts of the P~wave between the gold standard FEM approach and each of the BEM and IVC method quantified to 0.94 and 0.83 for fibrosis modeled as slow conducting tissue as well as 0.93 and 0.84 for the percolation modeling approach, respectively. In both simulation scenarios, especially the IVC method yielded too high ECGs amplitudes in the precordial leads and inaccurately captured atrial repolarization in the inferior leads II, III and aVF.

\section{Discussion}
\label{sec:discussion}
\subsection{Main Findings}
In this work, we compared atrial APD$_{90}$, LATs and ECGs computed with the bidomain, monodomain, reaction-eikonal and the eikonal propagation models as well as with the finite element, the boundary element and the infinite volume conductor forward calculation methods.

The largest deviations in LATs were observed between the bidomain simulations with Clerc and Roberts conductivity ratios. As the absolute LAT errors increase with the propagating wavefront, discrepancies in LATs can be traced back to more pronounced bath loading effects occurring in the Roberts conductivity settings. With a higher ratio between extracellular bulk and isotropic bath conductivities, the depolarization wave propagates faster in close vicinity to the interface between blood pool and endocardial wall leading to earlier LATs throughout the cardiac tissue. Due to the thin atrial wall, the bathloading effect is visible transmurally and thus leads to globally faster conduction velocities in the bidomain simulation with the Roberts conductivity setup. However, in this work, conductivities were tuned as described in section~\ref{sec:anatomicalModel} without a bath attached to one face of the strand meshes. Incorporating the bath already in the tuning process would have led to more similar results between the bidomain simulation results obtained with tuned Clerc and Roberts conductivities. 
This systematic underestimation of LATs also reflects in the ECG. The bidomain simulation with the Roberts conductivity settings yielded the smallest P~wave correlation to the bidomain ECGs with the Clerc conductivity ratios and markedly shorter P~wave duration. Also Sebastian \textit{et al.}~\cite{Sebastian-2008-ID17361} found that the choice of conductivity ratios in the intra- and extracellular domain as well as in longitudinal and transversal fiber direction had a marked effect on CV and LATs. Intra- and extracellular conductivity values were derived by Clerc and Roberts \textit{et al.} in animal experiments on specimen from excised trabecular cardiac bundles. Measuring intra- and extracellular current flow using microelectrodes allowed for a computation of the resistance and in turn the conductivity in longitudinal and transversal fiber direction in both, the extra- and intracellular space. Considering the complex and cumbersome in and ex vivo experiments to derive these parameters, fixed ratios between $\sigma_i$ and $\sigma_e$ along and perpendicular to the myocardial fiber orientation need to be assumed when personalizing computer models. As a consequence, the high uncertainty of the ratio between $\sigma_i$ and $\sigma_e$ which cannot be measured patient-specifically with reasonable efforts further justify the application of simplified models that do not involve uncertainties in non-measurable entities and only cause minor differences in LATs, ECGs and APD$_{90}$.

Among all investigated simplified model solutions, the monodomain model yielded the most accurate results regarding activation times, repolarization behavior and ECGs. However, explicit conductivity tuning for the monodomain model neither had a notable effect on LATs, nor APD$_{90}$, nor on the 12-lead ECG.

With mean and standard deviation of the absolute LAT differences to the bidomain results quantifying to 1.37$\pm$1.16\,ms and 1.43$\pm$1.17\,ms for the reaction-eikonal and the eikonal model, respectively, which differed only slightly due to numerical jitter. The distribution of LAT discrepancies to the bidomain results mapped on the atrial geometry was similar for the eikonal and the reaction-eikonal model. The LATs of the simplified propagation models were especially higher compared to the bidomain results in regions on the posterior left atrial wall. In these areas, different wavefronts collided causing an acceleration of the wave in the bidomain model, which is not captured in the (reaction-)eikonal model. Source-sink mismatch effects caused by convex wavefronts entailing conduction slowing in the bidomain model cause smaller LATs in the eikonal simulation results. This effect is especially visible in the area where Bachmann's bundle connects to the anterior wall of the left atrium, i.e. where a small source (Bachmann's bundle) meets a large sink (the left atrium). At the apex of the right atrial appendage, two convex wavefronts traversing the tissue from the lateral and the septal right atrial wall collide and cause eikonal LATs to be smaller than the ones resulting from the bidomain simulation. The P~waves computed with the reaction-eikonal and the eikonal source distribution showed similar correlation coefficients of 0.921 and 0.920 to the bidomain results. 
However, when evaluating repolarization dynamics, the reaction-eikonal model clearly led to more precise results. This reflects on the one side in smaller APD$_{90}$ discrepancies to the bidomain simulation results. The small APD$_{90}$ discrepancies between the monodomain and reaction-eikonal simulation results might have occurred due to differences in the activation pattern or a mismatch between the diffusion term and the $I_{foot}$ current in the case of curved wavefronts or wave collisions causing different AP upstrokes and amplitudes which subsequently lead to subtle APD changes. On the other hand, the reaction-eikonal model is capable of faithfully replicating both the P~wave as well as the atrial repolarization phase in the ECG, whereas with the eikonal model, only the P~wave highly resembles the bidomain results. Using precomputed AP templates to obtain the transmembrane voltage source distribution for the eikonal LAT results, APD$_{90}$ results were systematically smaller compared to the bidomain results in regular bulk tissue regions and systematically higher in fibrotic regions. The more precise representation of repolarization behavior in simulation results using the reaction-eikonal model is due to local APD balancing caused by the diffusion term. Consequently, also the repolarization signal in the ECG obtained with the source distribution derived from the eikonal results only showed a correlation coefficient of 0.62 to the bidomain ECG. 

ECGs calculated with the BEM highly resembled the ECGs obtained with the FEM. P~wave correlation coefficients to the FEM approach quantified to 0.94 and 0.93 for the simulation scenario with fibrosis modeled as slow conducting and non-conductive patches, respectively. In the former scenario, transmembrane voltages can be used as a source model for the forward calculation, whereas in the latter, volumetric sources such as primary impressed currents were necessary to model the effect of passive conduction barrier not contributing to the electrical source distribution in the heart. If the surface transmembrane voltages had been used as sources for the forward calculation in this case as well, an offset in the isoelectric line in the P~wave would have been induced. 
The infinite volume conductor method instead yielded more inaccurate ECG results. Especially in the septal and anterior leads, the ECG amplitudes were overestimated by a factor of $>$2 compared to the FEM results. On the one side, this observation can be traced back to the method's assumption that the atria are immersed in an infinite medium of a homogeneous conductivity, which does not allow considering a heterogeneous conductivity setup in the torso. On the other hand, the high ECG errors occurred predominantly in leads measured at electrode locations on the body surface in close proximity to the cardiac sources. Thus, neglecting the attenuating effect that the bounded torso volume conductor introduces causes a more pronounced effect on the resulting ECGs in V1-V3.

Simulations were run on a 16 core CPU machine (Intel Xeon Gold 6230, 2.1\,GHz). The full bidomain and the pseudo-bidomain simulation for a duration of 450\,ms were completed in 25 and 1.5 hours, respectively. Computation time for the reaction-eikonal setup was 1.4 hours on a 6 core machine.   
The computation of the transfer matrix for the BEM approach in the case of a heterogeneous torso volume conductor with seven surfaces bounding the atria, the torso and other organs took 2 hours on a 4 core CPU machine (Intel Core i5, 2.4\,GHz). The speed-up in computation times when using simplified propagation models is comparable to a ventricular setup. Computational performance improved by one order of magnitude when using the monodomain model~\cite{Bishop-2011-ID12010} and up to three orders of magnitude when using the eikonal or reaction-eikonal model~\cite{neic17, wallman12} compared to bidomain. 
Computational savings using the BEM approach are on the one hand due to the decreased problem complexity when discretizing the domain with surface instead of volume elements~\cite{potse09a}. On the other hand, coarser resolution meshes can be applied which is the key influencing factor for an improved computational efficiency over FEM~\cite{Mukherjee-1984-ID17366}.

\subsection{Related work}
In our work, we studied the differences in activation and repolarization times when using different propagation models in atrial electrophysiology, which, to the best of our knowledge, has not been done before in a comprehensive way. However, comparable studies have partly already been conducted for the ventricles and four-chamber heart models. Potse \textit{et al.}~\cite{Potse-2006-ID15641} found that activation using bidomain was 2\,\% faster compared to the monodomain approach for a complete cardiac cycle. Also in our study, the monodomain activation times were on average 1\,ms higher than those obtained from the bidomain simulation. 
Pashaei \textit{et al.}~\cite{Pashaei-0000-ID17310, Pashaei-2011-ID17309} as well as Wallman \textit{et al.}~\cite{wallman12} found that the differences in activation times are small for a ventricular simulation setup when comparing biophysically detailed approaches and the eikonal model. Neic \textit{et al.}~\cite{neic17} compared extracellular potential fields, electrograms and ECGs calculated with the reaction-eikonal and the bidomain model for the ventricles and concluded that the simplified model can replicate the gold standard results with high fidelity. Gassa \textit{et al.}~\cite{Gassa-2021-ID17311} investigated the suitability of an reaction-eikonal model to generate re-entrant activity on a bi-atrial geometry and succeeded in replicating the wave patterns resulting from a monodomain simulation. We have also recently shown that the eikonal-based models can produce activation times and ECGs resembling full bidomain simulation results with high fidelity in an atrial model without cellular remodeling placed in a homogeneous torso volume conductor~\cite{iHeart_nagel_2021}. Here, we extended the setup to heterogeneous scenarios covering cellular and conductivity heterogeneity in both the torso and the atria and observed similar results. This work confirms the findings from previous studies mainly conducted for ventricular simulation setups. 

Previous studies have also investigated the application of simplified forward calculation methods to computed ECGs. Schuler \textit{et al.}~\cite{Schuler-2019-ID12702} suggest the calculation of ECGs based on the BEM with coarse resolution surface meshes bounding the heart and the torso whereby parameters to blur the cardiac sources are optimized beforehand to avoid discontinuous wave propagation. In this way, they obtained body surface potentials in accurate accordance with the bidomain simulation results for a ventricular setup. However, one major drawback of the BEM approach is the impossibility of accounting for anisotropic conductivity in the myocardium~\cite{potse09a}. However, we found that P~wave correlation coefficients still quantified to $>$0.93 showing that the isotropic assumption yields similar ECGs compared to the bidomain results. For the infinite volume conductor method instead, not only the assumption of isotropic myocardial conductivities but also of a homogeneous torso volume conductor has to be made. Moreover, the simplified assumption that the atria is immersed in a medium of infinite spatial extent holds. Although the general P~wave morphology was preserved, the ECG still substantially differs regarding peak-to-peak amplitudes in the precordial leads and atrial repolarization in the inferior leads as it reflects in our results and was reported in previous work~\cite{unknown-0000-ID15784}.

\subsection{Limitations}
In this work, we investigated 4 different simulation scenarios comprising a healthy baseline case and three atrial models infiltrated with fibrosis, which was modeled either as slow conducting patches, non-conductive conduction barriers or ionic conductance rescaling. For the spatially distributed fibrotic areas (patchy and diffuse), none of the fibrosis remodeling scenarios had a marked effect on the ECG compared to the healthy baseline case. Ionic conductance rescaling, slow conducting fibrotic patches and percolation reflect in the ECG as a slight prolongation of the repolarization phase and an offset in the isoelectric line, a marginal prolongation of the P~wave and a decrease in peak-to-peak P~wave amplitudes, respectively. However, all these effects on the ECG are small and would show up in a more pronounced way if different fibrosis remodeling approaches were combined~\cite{Nagel-2021-ID16114}. However, we intentionally decided to investigate the effect of different propagation models and forward calculation methods in each of these simulation scenarios separately to shed light on which fibrosis remodeling aspects can be accurately captured by the simplified model solutions.

CVs were derived from the values reported in~\cite{loewe15b}. Based on them, conductivities were computed using \textit{tuneCV}~\cite{costa-cinc-2010} as described in section~\ref{sec:anatomicalModel} using strand meshes. However, no bath loading effects, mesh and wavefront curvature were considered when tuning the CVs, which might lead to mismatching CVs and conductivities assigned to different regions in the more complex atrial geometry. Adding a bath in the experiments set up for the tuning process, could lead to more similar LAT and ECG results between the bidomain results with Clerc and Roberts conductivities on the bi-atrial geometry. Moreover, performing the tuning with a bath attached to the strand geometries would lead to different monodomain conductivities for the setups without explicit conductivity tuning while the conductivity values in case of explicit conductivity tuning for the monodomain simulation would remain unchanged. Conductivities in transverse and longitudinal direction needed to be scaled by a factor of 54 and 12, respectively. The tuning procedure caused the original transversal vs. longitudinal conductivity ratios reported by Clerc and Roberts \textit{et al.} to change while keeping intra- vs. extracellular conductivity ratios constant.  

The average edge length of the atrial geometry was 523\,$\mu$m. To quantify the numerical error arising due to the mesh resolution, we conducted experiments on a 5\,cm$\times$2\,cm$\times$2.8\,mm block mesh with a resolution of 528\,$\mu$m and a refined resolution of 265\,$\mu$m by linearly subdividing the elements (see Fig.~2 in the supplementary material). Using the same numerical settings as for the experiments on the bi-atrial geometry which are summarized in section II in the supplementary material, we ran a simulation of a planar wave passing through an isthmus and then propagating with a curved wavefront (see Fig.~2 in the supplementary material). Conductivities were adjusted using \textit{tuneCV}~\cite{costa-cinc-2010} as described in section~\ref{anatomicalModel} to the CV in the regular atrial bulk tissue region. Maximum LAT differences between the experiments on the coarse and the fine mesh were 1.2\,ms. Considering the total activation time in the block of 43\,ms, the error introduced by the coarse mesh resolution was 2\%. The root mean squared errors between two APs resulting from the simulations on the coarse and the fine mesh were 0.0186\,mV and 0.0491\,mV for the two nodes marked in Fig. 2 in the supplementary material. When adding a fibrotic region to the block, the maximum absolute LAT error between the experiments on the fine and the coarse mesh was 1.2\,ms ($\sim$2\,\%) as well (see Fig. 3 in the supplementary material) for planar wave propagating along fiber direction. The latter is approximately also the case in our bi-atrial simulation setup where the depolarization wavefront traverses the elliptically shaped fibrotic patches growing predominantly in fiber direction. However, if a notable transverse wave propagation had to be represented, our chosen mesh resolution of 523$\mu$m would have been too coarse to capture the wave propagation at a velocity of 0.15m/s. Thus, the mesh resolution chosen for the atrial model in this study might introduce an error of 2\,\%, which is equivalent to an absolute LAT error of $\sim$2\,ms on the bi-atrial mesh. Due to the small root mean squared error between the APs of the coarse and the fine mesh, no additional discretization error affecting APD$_{90}$ is expected. 
\section{Conclusion}
The results presented here show that the eikonal model is capable of faithfully producing LATs and P~waves compared to full bidomain simulations with a reduction of computation times by a factor of up to three orders of magnitude. However, propagation models neglecting diffusion terms lack the fidelity in terms of repolarization as shown by APD$_{90}$ deviations. Thus, reaction-eikonal models are needed e.g. in cases where repolarization dynamics are of significant importance such as e.g. for re-entry mechanism studies.
ECGs calculated with the BEM accurately resemble the FEM results for both P~waves and the ECG in the repolarization phase. When computing ECGs with the infinite volume conductor method, the systematic overestimation of peak-to-peak P~wave amplitudes especially in the precordial leads should be taken into account when evaluating P~wave features.

\section*{Funding Statement}
This work was supported by the EMPIR programme co-financed by the participating states and from the European Union’s Horizon 2020 research and innovation programme under grant MedalCare 18HLT07. The authors acknowledge support by the Deutsche Forschungsgemeinschaft (DFG, German Research Foundation) under grant 2093/6-1 and  by the state of Baden-Württemberg through bwHPC. This project has received funding from the European Union’s Horizon 2020 research and innovation programme under the Marie Skłodowska-Curie grant agreement No.860974.

\bibliographystyle{plain}
\bibliography{bib}

\begin{thebibliography}{10}

\bibitem{Azzolin-2021-ID16653}
Luca Azzolin, Claudia Nagel, Deborah Nairn, Jorge Sanchez, Tianbao Zheng,
  Martin Eichenlaub, Amir Jadidi, Olaf Doessel, and Axel Loewe.
\newblock Automated framework for the augmentation of missing anatomical
  structures and generation of personalized atrial models from clinical data.
\newblock In {\em Computing in Cardiology Conference (CinC)}, 9 2021.

\bibitem{Azzolin-2021-ID16137}
Luca Azzolin, Steffen Schuler, Olaf Dössel, and Axel Loewe.
\newblock A reproducible protocol to assess arrhythmia vulnerability : Pacing
  at the end of the effective refractory period.
\newblock {\em Frontiers in Physiology}, 12:656411, 1 2021.

\bibitem{unknown-0000-ID15784}
Laura~R Bear, Leo~K Cheng, Ian~J LeGrice, Gregory~B Sands, Nigel~A Lever,
  David~J Paterson, and Bruce~H Smaill.
\newblock Forward problem of electrocardiography: is it solved?
\newblock {\em Circulation. Arrhythmia and electrophysiology}, 8(3):677--84, 6
  2015.

\bibitem{Bishop-2011-ID12010}
Martin~J Bishop and Gernot Plank.
\newblock Bidomain {ECG} simulations using an augmented monodomain model for
  the cardiac source.
\newblock {\em {IEEE} transactions on bio-medical engineering}, 58(8), 8 2011.

\bibitem{bishop2011bidomain}
Martin~J Bishop and Gernot Plank.
\newblock Bidomain ecg simulations using an augmented monodomain model for the
  cardiac source.
\newblock {\em IEEE transactions on biomedical engineering}, 58(8):2297--2307,
  2011.

\bibitem{Clerc-1976-ID12471}
L~Clerc.
\newblock Directional differences of impulse spread in trabecular muscle from
  mammalian heart.
\newblock {\em The Journal of Physiology}, 255(2):335--346, 1 1976.

\bibitem{costa-cinc-2010}
Caroline Costa, Elena Hoetzl, Bernardo Rocha, Anton Prassl, and Gernot Plank.
\newblock Automatic parameterization strategy for cardiac electrophysiology
  simulations.
\newblock {\em Computing in Cardiology (CinC)}, 2013.

\bibitem{Courtemanche-1998-ID17067}
Marc Courtemanche, Rafael~J. Ramirez, and Stanley Nattel.
\newblock Ionic mechanisms underlying human atrial action potential properties:
  insights from a mathematical model.
\newblock {\em American Journal of Physiology-Heart and Circulatory
  Physiology}, 275(1):H301--H321, 1998.

\bibitem{franzone2014mathematical}
Piero~Colli Franzone, Luca~Franco Pavarino, and Simone Scacchi.
\newblock {\em Mathematical cardiac electrophysiology}, volume~13.
\newblock Springer, 2014.

\bibitem{Gassa-2021-ID17311}
Narimane Gassa, Nejib Zemzemi, Cesare Corrado, and Yves Coudière.
\newblock Spiral waves generation using an eikonal-reaction cardiac
  electrophysiology model.
\newblock 12738:523--530, 1 2021.

\bibitem{geselowitz83}
D.~B. Geselowitz and T.~W. Miller.
\newblock A bidomain model for anisotropic cardiac muscle.
\newblock {\em Annals of Biomedical Engineering}, 11(3-4):191--206, 1 1983.

\bibitem{Gillette-2021-ID16142}
Karli Gillette, Matthias~A.F. Gsell, Anton~J. Prassl, Elias Karabelas, Ursula
  Reiter, Gert Reiter, Thomas Grandits, Christian Peyer, Darko Štern, Martin
  Urschler, Jason~D. Bayer, Christoph~M. Augustin, Aurel Neic, Thomas Pock,
  Edward~J. Vigmond, and Gernot Plank.
\newblock A framework for the generation of digital twins of cardiac
  electrophysiology from clinical 12-leads {ECG}s.
\newblock {\em Medical Image Analysis}, page 102080, 1 2021.

\bibitem{Higuchi-2018-ID11688}
Koji Higuchi, Joshua Cates, Gregory Gardner, Alan Morris, Nathan~S. Burgon,
  Nazem Akoum, and Nassir~F. Marrouche.
\newblock The spatial distribution of late gadolinium enhancement of left
  atrial magnetic resonance imaging in patients with atrial fibrillation.
\newblock {\em {JACC}: Clinical Electrophysiology}, 4(1):49--58, 1 2018.

\bibitem{Jakob-2015-ID12648}
Wenzel Jakob, Marco Tarini, Daniele Panozzo, and Olga Sorkine-Hornung.
\newblock Instant field-aligned meshes.
\newblock {\em {ACM} Transactions on Graphics}, 34(6):1--15, 1 2015.

\bibitem{keener2009mathematical}
JP~Keener and James Sneyd.
\newblock Mathematical physiology 1: Cellular physiology, 2009.

\bibitem{krueger13}
M.~W. Krueger, A.~Dorn, D.~U.~J. Keller, F.~Holmqvist, J.~Carlson, P.~G.
  Platonov, K.~S. Rhode, R.~Razavi, G.~Seemann, and O.~Dössel.
\newblock In-silico modeling of atrial repolarization in normal and atrial
  fibrillation remodeled state.
\newblock {\em Medical \& Biological Engineering \& Computing},
  51(10):1105--1119, 1 2013.

\bibitem{krueger13b}
M.~W. Krueger, G.~Seemann, K.~Rhode, D.~U.~J. Keller, C.~Schilling, A.~Arujuna,
  J.~Gill, M.~D. O'Neill, R.~Razavi, and O.~Dössel.
\newblock Personalization of atrial anatomy and electrophysiology as a basis
  for clinical modeling of radio-frequency ablation of atrial fibrillation.
\newblock {\em {IEEE} Transactions on Medical Imaging}, 32(1):73--84, 1 2013.

\bibitem{krueger13e}
M.~W. Krüger.
\newblock {\em Personalized Multi-Scale Modeling of the Atria :
  Heterogeneities, Fiber Architecture, Hemodialysis and Ablation Therapy}.
\newblock PhD thesis, 1 2013.

\bibitem{loewe15b}
A.~Loewe, M.~W. Krueger, P.~G. Platonov, F.~Holmqvist, O.~Dössel, and
  G.~Seemann.
\newblock Left and right atrial contribution to the p-wave in realistic
  computational models.
\newblock {\em Functional Imaging and Modeling of the Heart 2015, Lecture Notes
  in Computer Science}, (9126):439--447, 1 2015.

\bibitem{Luongo-2022-ID17367}
Giorgio Luongo, Gaetano Vacanti, Vincent Nitzke, Deborah Nairn, Claudia Nagel,
  Diba Kabiri, Tiago~P Almeida, Diogo~C Soriano, Massimo~W Rivolta,
  Ghulam~André Ng, Olaf Dössel, Armin Luik, Roberto Sassi, Claus Schmitt, and
  Axel Loewe.
\newblock Hybrid machine learning to localize atrial flutter substrates using
  the surface 12-lead electrocardiogram.
\newblock {\em {EP} Europace}, 1 2022.

\bibitem{Mukherjee-1984-ID17366}
Subrata Mukherjee and Mahesh Morjaria.
\newblock On the efficiency and accuracy of the boundary element method and the
  finite element method.
\newblock {\em International Journal for Numerical Methods in Engineering},
  20:515--522, 3 1984.

\bibitem{iHeart_nagel_2021}
Claudia Nagel, Cristian~Alberto Barrios~Espinosa, Karli Gillette, Matthias
  Gsell, Gernot Plank, Olaf Dössel, and Axel Loewe.
\newblock Comparison of source models and forward calculation methods for
  atrial electrophysiology regarding activation times and electrocardiograms.
\newblock In {\em i{HEART} Congress – Modelling the Cardiac Function}, 2021.

\bibitem{Nagel-2021-ID16114}
Claudia Nagel, Giorgio Luongo, Luca Azzolin, Steffen Schuler, Olaf Dössel, and
  Axel Loewe.
\newblock Non-invasive and quantitative estimation of left atrial fibrosis
  based on p waves of the 12-lead {ECG}-a large-scale computational study
  covering anatomical variability.
\newblock {\em Journal of Clinical Medicine}, 10(8), 4 2021.

\bibitem{neic17}
A.~Neic, F.~O. Campos, A.~J. Prassl, S.~A. Niederer, M.~J. Bishop, E.~J.
  Vigmond, and G.~Plank.
\newblock Efficient computation of electrograms and {ECG}s in human whole heart
  simulations using a reaction-eikonal model.
\newblock {\em Journal of Computational Physics}, 346:191--211, 1 2017.

\bibitem{neic2017efficient}
Aurel Neic, Fernando~O Campos, Anton~J Prassl, Steven~A Niederer, Martin~J
  Bishop, Edward~J Vigmond, and Gernot Plank.
\newblock Efficient computation of electrograms and ecgs in human whole heart
  simulations using a reaction-eikonal model.
\newblock {\em Journal of computational physics}, 346:191--211, 2017.

\bibitem{Neic-2020-ID13697}
Aurel Neic, Matthias~A.F. Gsell, Elias Karabelas, Anton~J. Prassl, and Gernot
  Plank.
\newblock Automating image-based mesh generation and manipulation tasks in
  cardiac modeling workflows using meshtool.
\newblock {\em SoftwareX}, 11:100454, 1 2020.

\bibitem{niederer2011verification}
Steven~A Niederer, Eric Kerfoot, Alan~P Benson, Miguel~O Bernabeu, Olivier
  Bernus, Chris Bradley, Elizabeth~M Cherry, Richard Clayton, Flavio~H Fenton,
  Alan Garny, et~al.
\newblock Verification of cardiac tissue electrophysiology simulators using an
  n-version benchmark.
\newblock {\em Philosophical Transactions of the Royal Society A: Mathematical,
  Physical and Engineering Sciences}, 369(1954):4331--4351, 2011.

\bibitem{Pashaei-0000-ID17310}
A.~Pashaei, D.~Romero, R.~Sebastian, O.~Camara, and A.~F. Frangi.
\newblock Comparison of phenomenological and biophysical cardiac models coupled
  with heterogenous structures for prediction of electrical activation
  sequence.
\newblock 2010.

\bibitem{Pashaei-2011-ID17309}
A.~Pashaei, D.~Romero, R.~Sebastian, O.~Camara, and A.~F. Frangi.
\newblock Fast multiscale modeling of cardiac electrophysiology including
  purkinje system.
\newblock {\em {IEEE} Transactions on Biomedical Engineering},
  58(10):2956--2960, 1 2011.

\bibitem{pernod2011multi}
Erik Pernod, Maxime Sermesant, Ender Konukoglu, Jatin Relan, Herv{\'e}
  Delingette, and Nicholas Ayache.
\newblock A multi-front eikonal model of cardiac electrophysiology for
  interactive simulation of radio-frequency ablation.
\newblock {\em Computers \& Graphics}, 35(2):431--440, 2011.

\bibitem{Plank-2021-ID15953}
Gernot Plank, Axel Loewe, Aurel Neic, Christoph Augustin, Yung-Lin Huang,
  Matthias A~F Gsell, Elias Karabelas, Mark Nothstein, Anton~J Prassl, Jorge
  Sánchez, Gunnar Seemann, and Edward~J Vigmond.
\newblock The open{CARP} simulation environment for cardiac electrophysiology.
\newblock {\em Computer methods and Programs in Biomedicine}, 208:106223, 6
  2021.

\bibitem{potse09a}
M.~Potse, B.~Dube, and A.~Vinet.
\newblock Cardiac anisotropy in boundary-element models for the
  electrocardiogram.
\newblock {\em Medical \& Biological Engineering \& Computing}, 47(7):719--729,
  1 2009.

\bibitem{Potse-2006-ID15641}
Mark Potse, Bruno Dubé, Alain Vinet, and René Cardinal.
\newblock A comparison of monodomain and bidomain propagation models for the
  human heart.
\newblock {\em Conference proceedings : ... Annual International Conference of
  the {IEEE} Engineering in Medicine and Biology Society. {IEEE} Engineering in
  Medicine and Biology Society. Annual Conference}, 2006:3895--8, 1 2006.

\bibitem{pullan2002finite}
Andrew~J Pullan, Karl~A Tomlinson, and Peter~J Hunter.
\newblock A finite element method for an eikonal equation model of myocardial
  excitation wavefront propagation.
\newblock {\em SIAM Journal on Applied Mathematics}, 63(1):324--350, 2002.

\bibitem{reinke14}
A.~Reinke, D.~Potyagaylo, W.~H.~W. Schulze, and O.~Dössel.
\newblock Geometrical model and corresponding conductivities for solving the
  inverse problem of {ECG}.
\newblock In {\em Biomedizinische Technik / Biomedical Engineering}, volume~59,
  pages 937--940, 1 2014.

\bibitem{roberts82}
D.~E. Roberts and A.~M. Scher.
\newblock Effect of tissue anisotropy on extracellular potential fields in
  canine myocardium in situ.
\newblock {\em Circ. Res.}, 50:342--351, 1 1982.

\bibitem{Roney-2016-ID12208}
Caroline~H Roney, Jason~D Bayer, Sohail Zahid, Marianna Meo, Patrick M~J Boyle,
  Natalia~A Trayanova, Michel Haïssaguerre, Rémi Dubois, Hubert Cochet, and
  Edward~J Vigmond.
\newblock Modelling methodology of atrial fibrosis affects rotor dynamics and
  electrograms.
\newblock {\em {EP} Europace}, 18(suppl 4):iv146--iv155, 12 2016.

\bibitem{Schuler-2019-ID12702}
Steffen Schuler, Jess~D. Tate, Thom~F. Oostendorp, Robert~S. MacLeod, and Olaf
  Dössel.
\newblock Spatial downsampling of surface sources in the forward problem of
  electrocardiography.
\newblock In Yves Coudière, Valéry Ozenne, Edward Vigmond, and Nejib Zemzemi,
  editors, {\em Functional Imaging and Modeling of the Heart}, volume 11504 of
  {\em Lecture Notes in Computer Science}, pages 29--36. Springer International
  Publishing, 5 2019.

\bibitem{Sebastian-2008-ID17361}
Rafael Sebastian, Sebastian Ordas, Gernot Plank, Blanca Rodriguez, Edward~J.
  Vigmond, and Alejandro~F. Frangi.
\newblock Assessing influence of conductivity in heart modelling with the aim
  of studying cardiovascular diseases.
\newblock {\em Medical Imaging 2008: Physiology}, 6916:691627, 3 2008.

\bibitem{stenroos07}
M.~Stenroos, V.~Mäntynen, and J.~Nenonen.
\newblock A matlab library for solving quasi-static volume conduction problems
  using the boundary element method.
\newblock {\em Computer methods and programs in biomedicine}, 88(3):256--263, 1
  2007.

\bibitem{tung1978bi}
Leslie Tung.
\newblock {\em A bi-domain model for describing ischemic myocardial dc
  potentials.}
\newblock PhD thesis, Massachusetts Institute of Technology, 1978.

\bibitem{vigmond16}
E.~Vigmond, A.~Pashaei, S.~Amraoui, H.~Cochet, and M.~Hassaguerre.
\newblock Percolation as a mechanism to explain atrial fractionated
  electrograms and reentry in a fibrosis model based on imaging data.
\newblock {\em Heart Rhythm}, 13(7):1536--1543, 1 2016.

\bibitem{Vigmond-2003-ID14313}
Edward~J Vigmond, Matt Hughes, G~Plank, and L~Joshua Leon.
\newblock Computational tools for modeling electrical activity in cardiac
  tissue.
\newblock {\em Journal of electrocardiology}, 36 Suppl:69--74, 1 2003.

\bibitem{vigmond2008solvers}
EJ~Vigmond, R~Weber Dos~Santos, AJ~Prassl, M~Deo, and G~Plank.
\newblock Solvers for the cardiac bidomain equations.
\newblock {\em Progress in biophysics and molecular biology}, 96(1-3):3--18,
  2008.

\bibitem{wallman12}
M.~Wallman, N.~P. Smith, and B.~Rodriguez.
\newblock A comparative study of graph-based, eikonal, and monodomain
  simulations for the estimation of cardiac activation times.
\newblock {\em {IEEE} Transactions on Biomedical Engineering},
  59(6):1739--1748, 1 2012.

\bibitem{Zheng-2021-ID17236}
Tianbao Zheng, Luca Azzolin, Jorge Sánchez, Olaf Dössel, and Axel Loewe.
\newblock An automate pipeline for generating fiber orientation and region
  annotation in patient specific atrial models.
\newblock In {\em Current Directions in Biomedical Engineering}, volume~7,
  pages 136--139, 1 2021.

\end{thebibliography}

\pagebreak
\begin{center}
\textbf{\large Supplemental Material}
\end{center}
\setcounter{equation}{0}
\setcounter{figure}{0}
\setcounter{table}{0}
\makeatletter
\renewcommand{\theequation}{S\arabic{equation}}
\renewcommand{\thefigure}{S\arabic{figure}}
\renewcommand{\bibnumfmt}[1]{[S#1]}
\renewcommand{\citenumfont}[1]{S#1}
\section*{Conduction Velocity and Conductivity Settings}
\begin{figure*}[!h]
\centerline{\includegraphics[width = \textwidth]{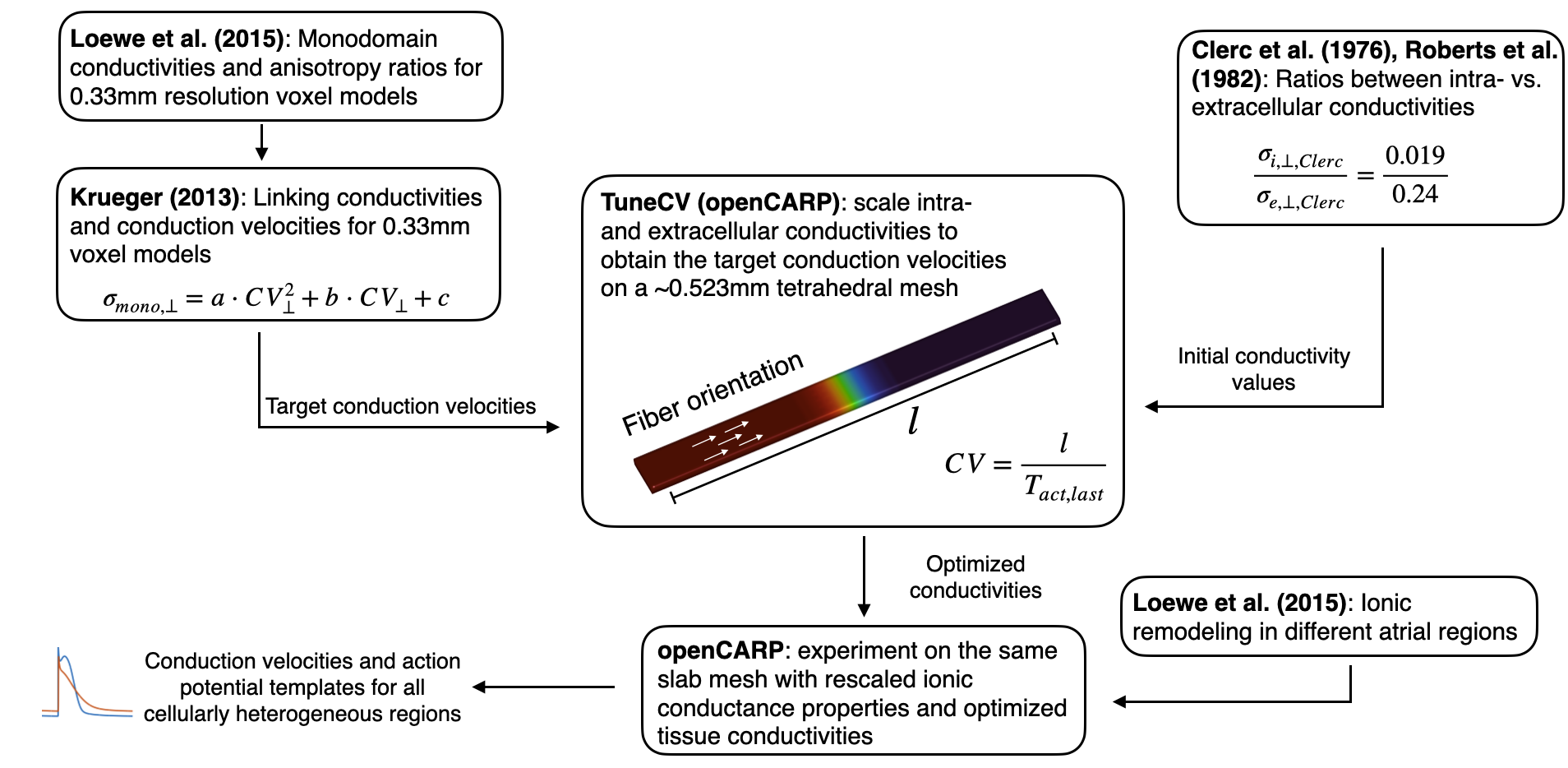}}
\caption{Workflow for tuning conductivities and conduction velocities with \textit{tuneCV}. The monodomain conductivities reported in Loewe et al. (2015) were transfered in conduction velocities using the formulas reported by Krueger (2013). On a slab mesh with a resolution corresponding to the average edge length of the regions in the bi-atrial geometry, the initial intra- and extracellular conductivities as reported by Clerc et al. as well as by Roberts et al. were assigned. In an iterative optimization process, these conductivities were linearly scaled until the target conduction velocity was reached. In an openCARP experiment, action potential templates were computed using the ionic remodeling settings reported by Loewe et al. (2015) and the optimized conductivity settings on the same slab mesh. }
\label{tuneCV_workflow}
\end{figure*}

\clearpage
\thispagestyle{empty}
\begin{landscape}
\begin{table}[!htb]
\caption{Conductivities calculated with the intra- and extracellular conductivity ratios reported by Clerc et al. and Roberts et al. as well as monodomain conductivities with and without explicit conductivity tuning. The monodomain conductivities in the latter case were obtained by computing half of the harmonic mean tensor from the bidomain conductivities. }
\begin{tiny}
\begin{center}
    
\begin{tabular}{l|l|llll|llll|ll|ll}
\textbf{Atrial region}      & \textbf{res ($\mu m$)} & \multicolumn{4}{c|}{\textbf{Roberts conductivities}}                                    & \multicolumn{4}{c|}{\textbf{Clerc conductivities}}                                        & \multicolumn{2}{c}{\textbf{Monodomain conductivities}} & \multicolumn{2}{c}{\textbf{Monodomain conductivities, no tuning}} \\ \hline
\textbf{}                   & \textbf{}              & $\sigma_i^{\parallel}$ & $\sigma_e^{\parallel}$ & $\sigma_i^{\perp}$ & $\sigma_e^{\perp}$ & $\sigma_i^{\parallel}$ & $\sigma_e^{\parallel}$ & $\sigma_i^{\perp}$ & $\sigma_e^{\perp}$ & $\sigma_m^{\parallel}$       & $\sigma_m^{\perp}$  & $\sigma_m^{\parallel}$       & $\sigma_m^{\perp}$   \\ \hline
Regular atrial myocardium   & 523                    & 2.9135             & 1.0283             & 0.3526                 & 0.4702                 & 1.0189             & 3.7160             & 0.2197                 & 2.7749                 & 0.4018 & 0.1018 & 0.3998 & 0.1018                    \\
Atrial Appendages           & 507                    & 2.9135             & 1.0283             & 0.3526                 & 0.4702                 & 1.0189             & 3.7160             & 0.2197                 & 2.7749                 & 0.4018 & 0.1018 & 0.3998 & 0.1018                      \\
Atrio-ventricular rings     & 501                    & 2.9135             & 1.0283             & 0.3526                 & 0.4702                 & 1.0189             & 3.7160             & 0.2197                 & 2.7749                 & 0.4018 & 0.1018 & 0.3998 & 0.1018                      \\
Crista terminalis           & 534                    & 5.3727             & 1.8962             & 0.3557                 & 0.4743                 & 1.8784             & 6.8507             & 0.2216                 & 2.7990                 & 0.7407 & 0.1028 & 0.7371 & 0.1027                     \\
Pectinate muscles           & 510                    & 5.7922             & 2.0443             & 0.2383                 & 0.3177                 & 2.0239             & 7.3814             & 0.1481                 & 1.8702                 & 0.7982 & 0.0687 & 0.7942 & 0.0686                     \\
Bachmann’s bundle           & 530                    & 8.8775             & 3.1332             & 0.4090                 & 0.5454                 & 3.0979             & 11.2980            & 0.2550                 & 3.2214                 & 1.2215 & 0.1182 & 1.2156 & 0.1181                      \\
Inferior isthmus            & 525                    & 0.6573             & 0.2320             & 0.3070                 & 0.4093                 & 0.2250             & 0.8207             & 0.1911                 & 2.4142                 & 0.0886 & 0.0886 & 0.0883 & 0.0885                     \\
Fibrosis (non conductive)   & 523                    & 0                  & 0                  & 0                      & 0                      & 0                  & 0                  & 0                      & 0                      & 0 & 0 & 0 & 0                           \\
Fibrosis (slow conducting)  & 523                    & 0.8208             & 0.2897             & 0.0466                 & 0.0621                 & 0.2823             & 1.0296             & 0.0288                 & 0.3644                 & 0.1112 & 0.0133 & 0.1108 & 0.0133                      \\
Fibrosis (ionic remodeling) & 523                    & 2.9135             & 1.0283             & 0.3526                 & 0.4702                 & 1.0189             & 3.7160             & 0.2197                 & 2.7749                 & 0.4018 & 0.1018 & 0.3998 & 0.1018                    
\end{tabular}
\label{tab:conductivities}
\end{center}
\end{tiny}
\end{table}
\end{landscape}
\clearpage

\begin{table*}[!h]
\caption{Scaling factors for ion channel conductances with respect to the baseline Courtemanche et al. cell model and resulting conduction velocities in different atrial regions.}
\begin{footnotesize}
\begin{center}

\begin{tabular}{l|llll|ll}
\textbf{Atrial region}      & \multicolumn{4}{c|}{\textbf{Ionic heterogeneity}} & \multicolumn{2}{c}{\textbf{Conduction velocity (m/s)}} \\ \hline &&&&&\\
\textbf{}                   & $g_{to}$  & $g_{CaL}$  & $g_{Na}$  & $g_{K1}$  & $CV^{\perp}$             & $CV^{\parallel}$            \\ &&&&&\\\hline

Regular atrial myocardium   &           &            &           &           & 0.5905                   & 1.2465                      \\
Atrial appendages           & 0.68      & 1.06       &           &           & 0.5950                   & 1.2467                      \\
Atrio-ventricular rings     & 1.53      & 0.67       &           &           & 0.5965                   & 1.2456                      \\
Crista terminalis           &        & 1.67       &           &           & 0.5911                   & 1.6839                      \\
Pectinate muscles           &           &            &           &           & 0.4612                   & 1.7435                      \\
Bachmann’s bundle           &           &            &           &           & 0.6450                   & 2.1511                      \\
Inferior isthmus            &           &            &           &           & 0.5402                   & 0.5402                      \\
Fibrosis (non conductive)   &           &            &           &           & 0                        & 0                           \\
Fibrosis (slow conducting)  &           &            &           &           & 0.1181                   & 0.6232                      \\
Fibrosis (ionic remodeling) &           & 0.5        & 0.6       & 0.5       & 0.4812                   & 1.0063                     
\end{tabular}

\label{tab:CV}
\end{center}
\end{footnotesize}
\end{table*}
\FloatBarrier
\section*{Numerical Simulation Settings}
Unless stated otherwise in the list below, the default numerical settings defined in openCARP (e.g. convergence criteria for PDE solver, surface-to-volume ratio) were applied to all experiments. Adapted values were chosen for: 
\begin{itemize}
\item output time resolution: $timedt = 1$\,ms
\item time step size to solve the numeric equations: $dt = 25\,\mu$s
\item mass lumping: \textit{false}
\end{itemize}
\FloatBarrier
\clearpage
\section*{Mesh Resolution Impact}
\begin{figure*}[!h]
\centerline{\includegraphics[width = \textwidth]{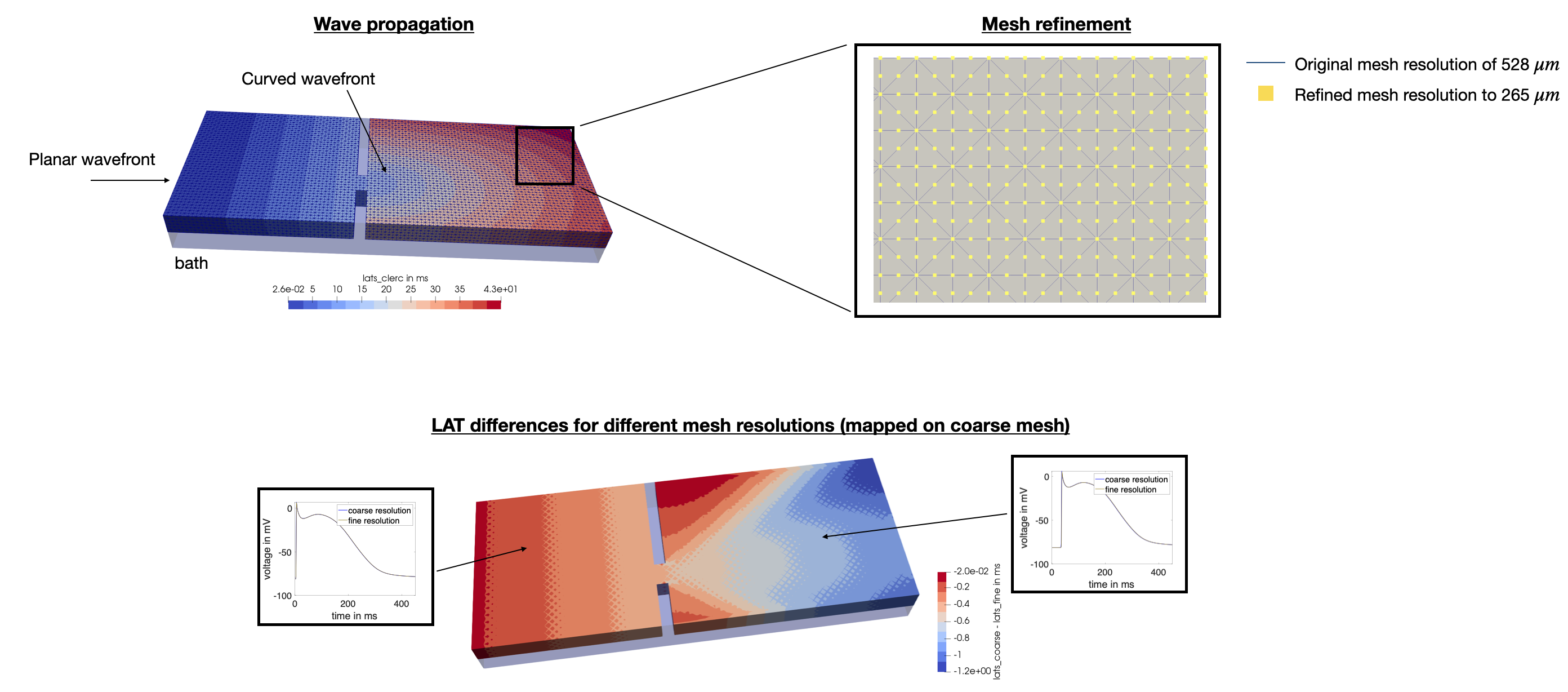}}
\caption{Experiment for quantifying the error in LATs due to a coarser mesh resolution in the bidomain model. The top panel shows the 5\,cm$\times$2\,cm$\times$2.8\,mm block mesh with a resolution of 528\,$\mu$m. The mesh resolution on the same geometry was refined to 265\,$\mu$m by linearly subdividing the tetrahedral elements (top right panel). The excitation was initiated by pacing from the left side of the block. When the planar wave passes through the isthmus, it propagates with a curved wavefront onwards. The bottom panel shows the signed LAT differences between the nodes in the coarse and the fine mesh. Two action potentials at nodes on the right and the left end of the block are visualized for the fine and the coarse resolution mesh.}
\label{meshresolution_baseline}
\end{figure*}

\begin{figure*}[!h]
\centerline{\includegraphics[width = \textwidth]{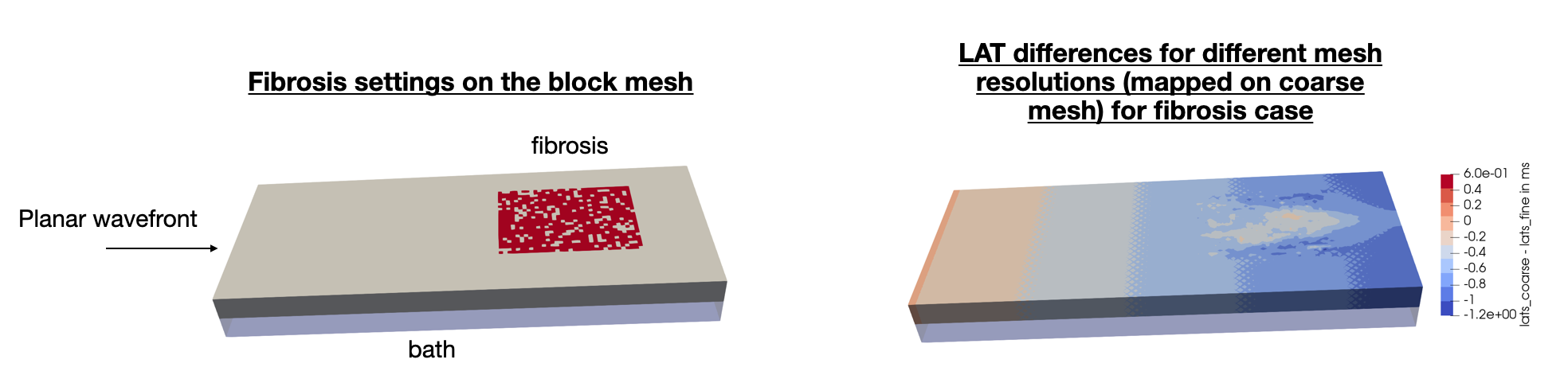}}
\caption{Experiment for quantifying the error in LATs due to a coarser mesh resolution in the bidomain model in case of fibrotically infiltrated tissue. The left panel shows the model setup with a fibrotic patch of 80\,\% density and a planar wave propagating along the block. The right panel shows the color-coded differences in LATs for the coarse and the fine mesh.}
\label{meshresolution_fibrosis}
\end{figure*}
\FloatBarrier
\clearpage
\section*{Effect of Propagation Models on LATs in Different Simulation Scenarios}
\begin{figure*}[!h]
\centerline{\includegraphics[width = \textwidth]{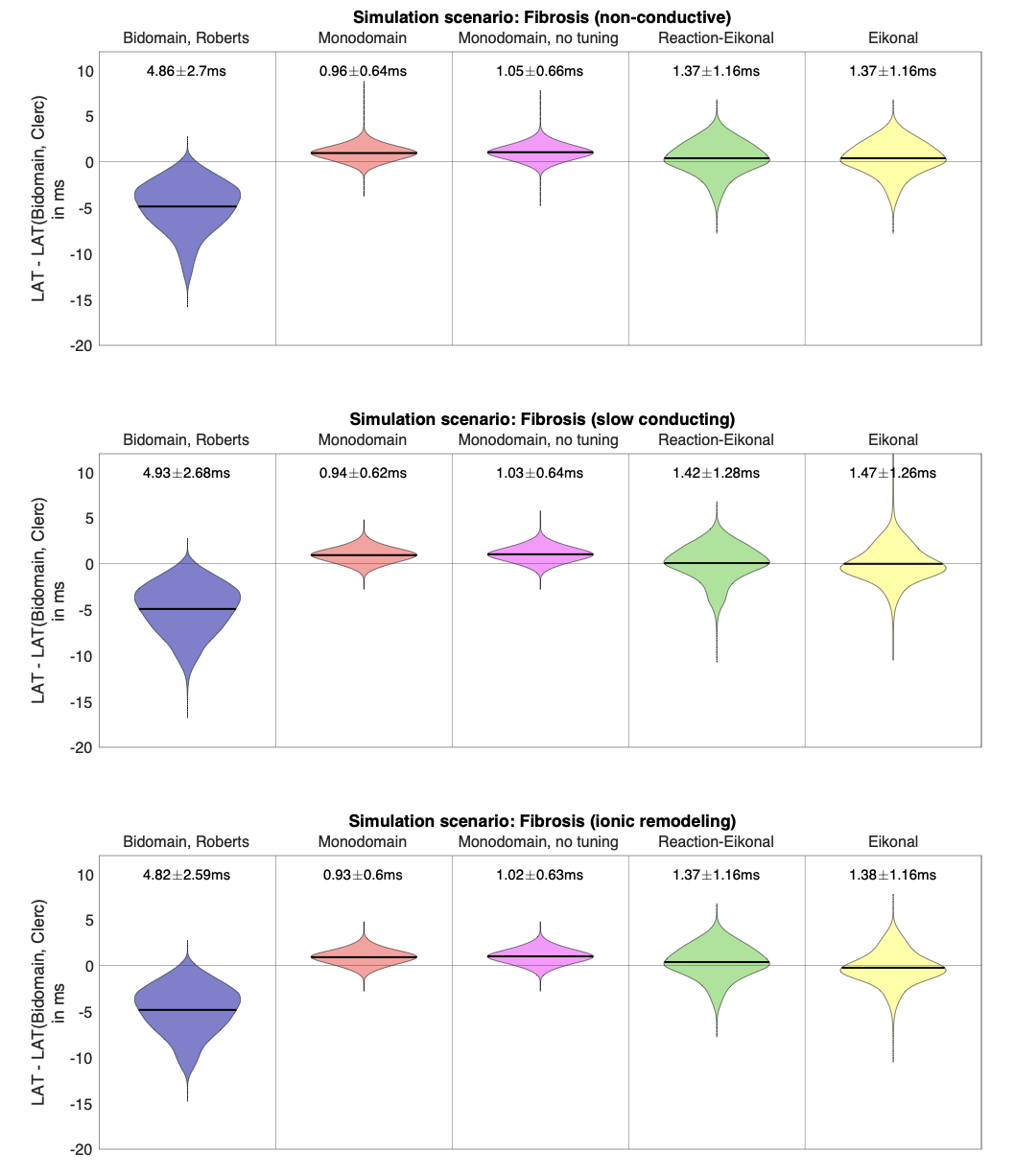}}
\caption{LAT differences between different propagation models and the bidomain simulation results obtained with the Clerc conductivities. The top row shows the mean $\pm$ standard deviation of the absolute differences to the bidomain LATs. }
\label{lat_pd_complete}
\end{figure*}

\begin{figure*}[!h]
\centerline{\includegraphics[width = \textwidth]{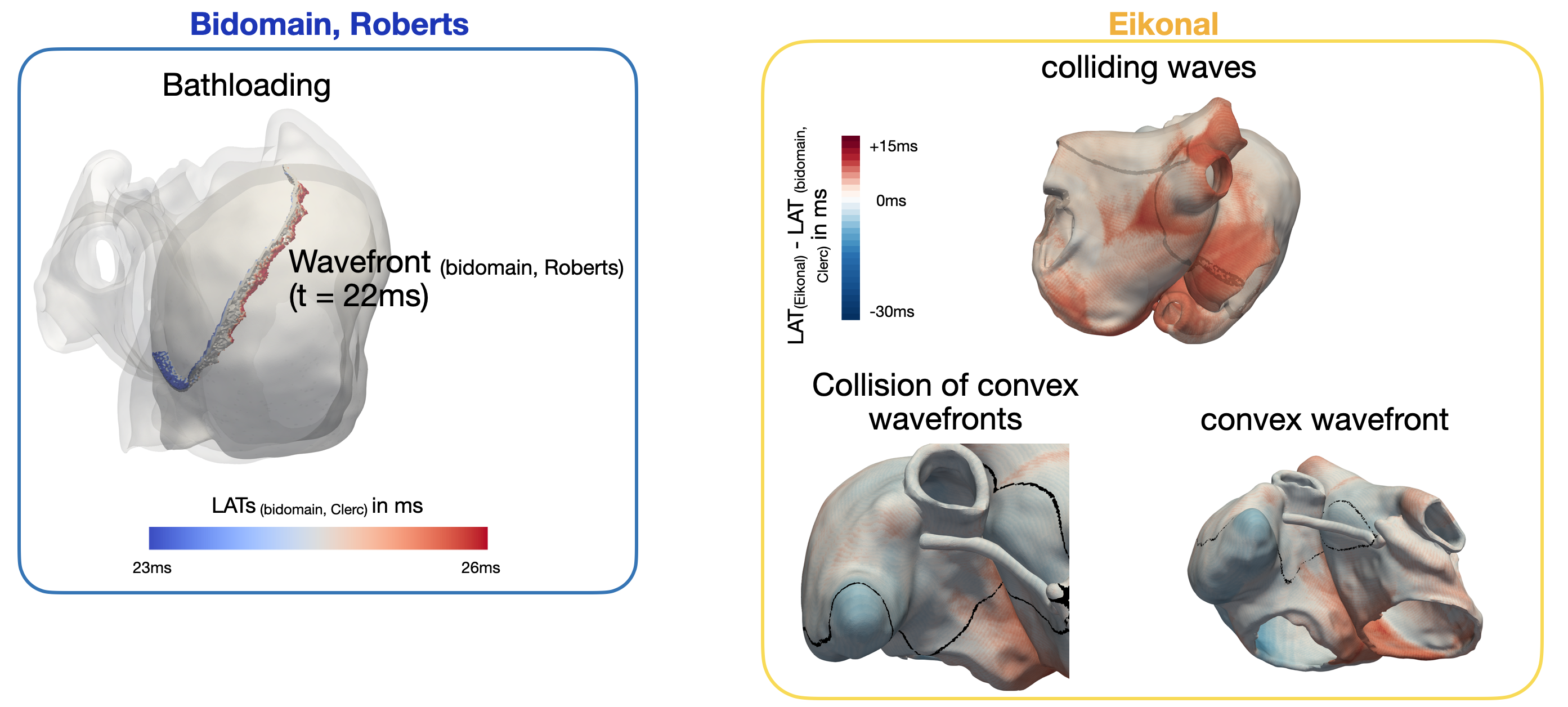}}
\caption{Visualization of LAT discrepancies between the bidomain results and other propagation models. The right panel shows the wavefront position obtained with the bidomain simulation ran with Roberts conductivities at t=22\,ms. The more pronounced bathloading effects for the Roberts conductivity ratios becomes visible by color coding the activation times obtained with the bidomain simulation and Clerc conductivity settings. In this case, nodes along the wavefront at the endocardium in close proximity to the interface between blood pool and myocardial tissue were excited at a later time than the ones at the epicardium. The right panel shows the signed LAT differences between the eikonal and the bidomain simulation. The top panel shows the atrial geometry from the posterior view where wavefronts collide and cause an acceleration of the wavefront in the bidomain, but not in the eikonal model. The bottom panel shows the effect of convex wavefronts at the right atrial appendage as well as at the connection between the Bachmann's bundle and the left atrium.   }
\label{explanation_lats}
\end{figure*}
\FloatBarrier
\clearpage
\section*{Effect of Propagation Models on APD90 in Different Simulation Scenarios}
\begin{figure*}[!h]
\centerline{\includegraphics[width = \textwidth]{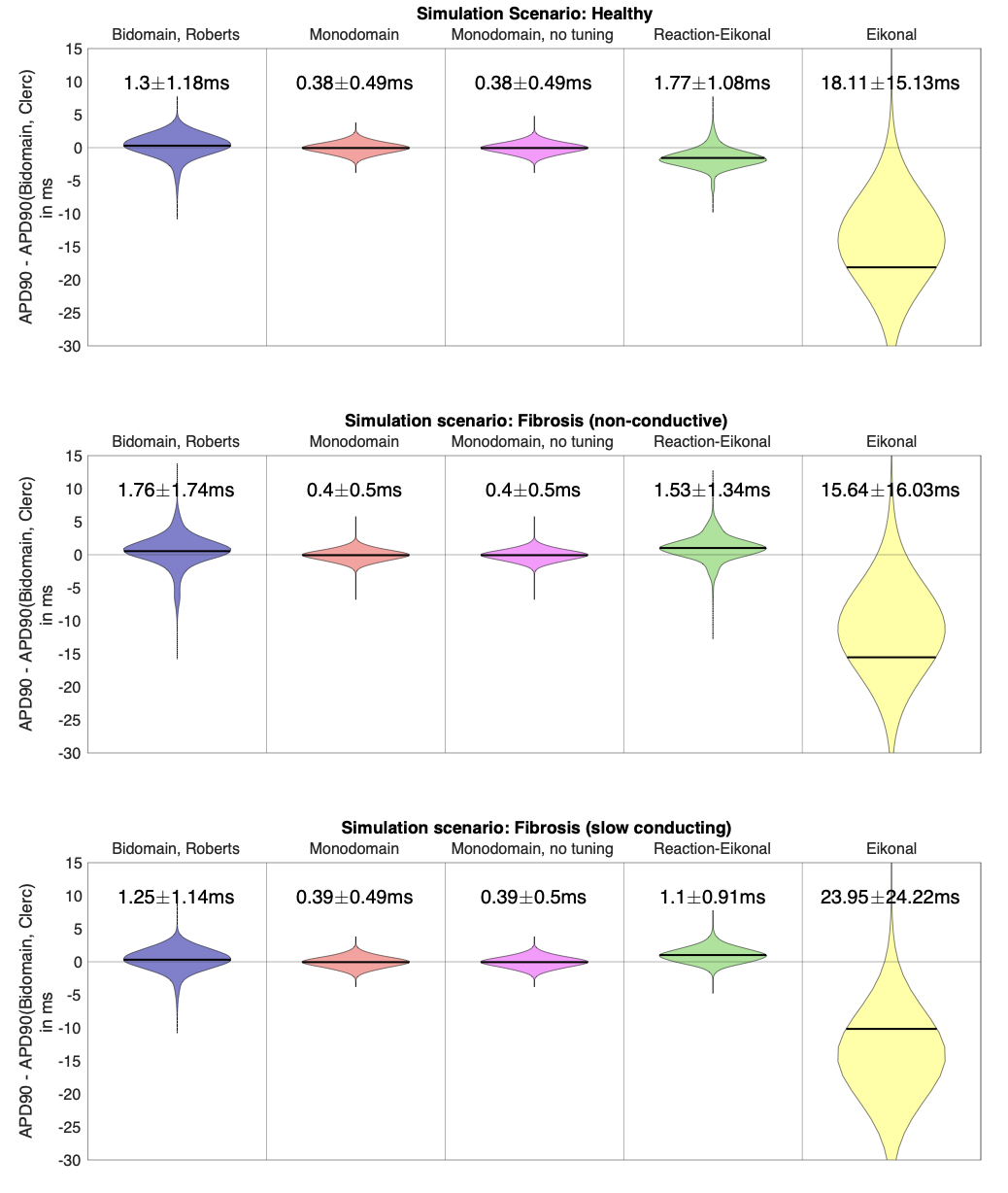}}
\caption{APD$_{90}$ differences between different propagation models and the bidomain simulation results obtained with the Clerc conductivities. The top row shows the mean $\pm$ standard deviation of the absolute differences to the bidomain APD$_90$ results.}
\label{apd90results}
\end{figure*}
\FloatBarrier
\clearpage
\section*{Effect of Propagation Models on ECGs and ECG Features in Different Simulation Scenarios}
\begin{figure*}[!h]
\centerline{\includegraphics[width = \textwidth]{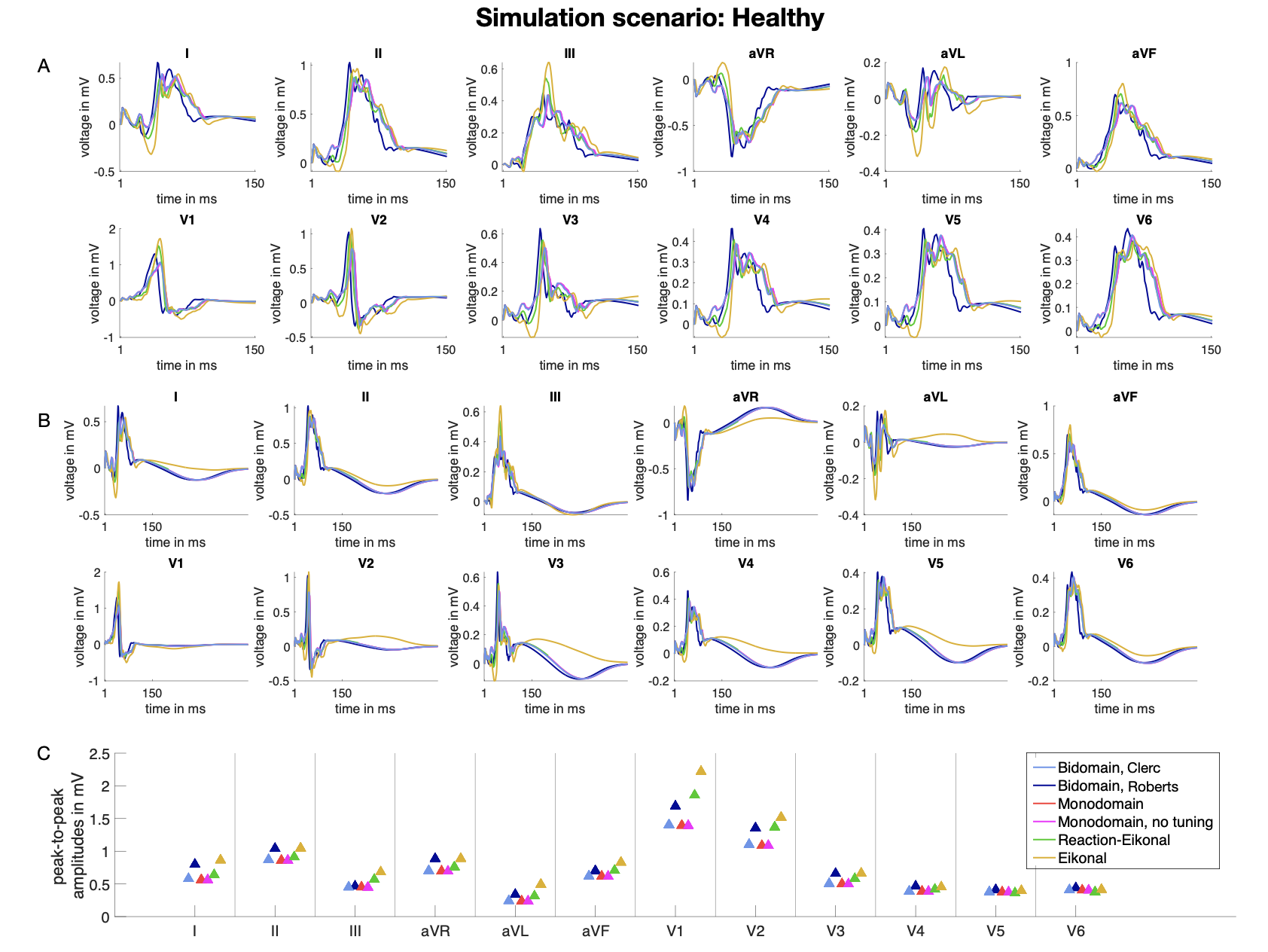}}
\caption{ECGs for different propagation models using the same forward calculation method (BEM) in the Interval [0, 150]\,ms (A) and [0, 650]\,ms (B) in the healthy case. Panel C shows the peak-to-peak amplitude features resulting for each propagation model. }
\label{ecg_pd_healthy}
\end{figure*}

\begin{figure*}[!h]
\centerline{\includegraphics[width = \textwidth]{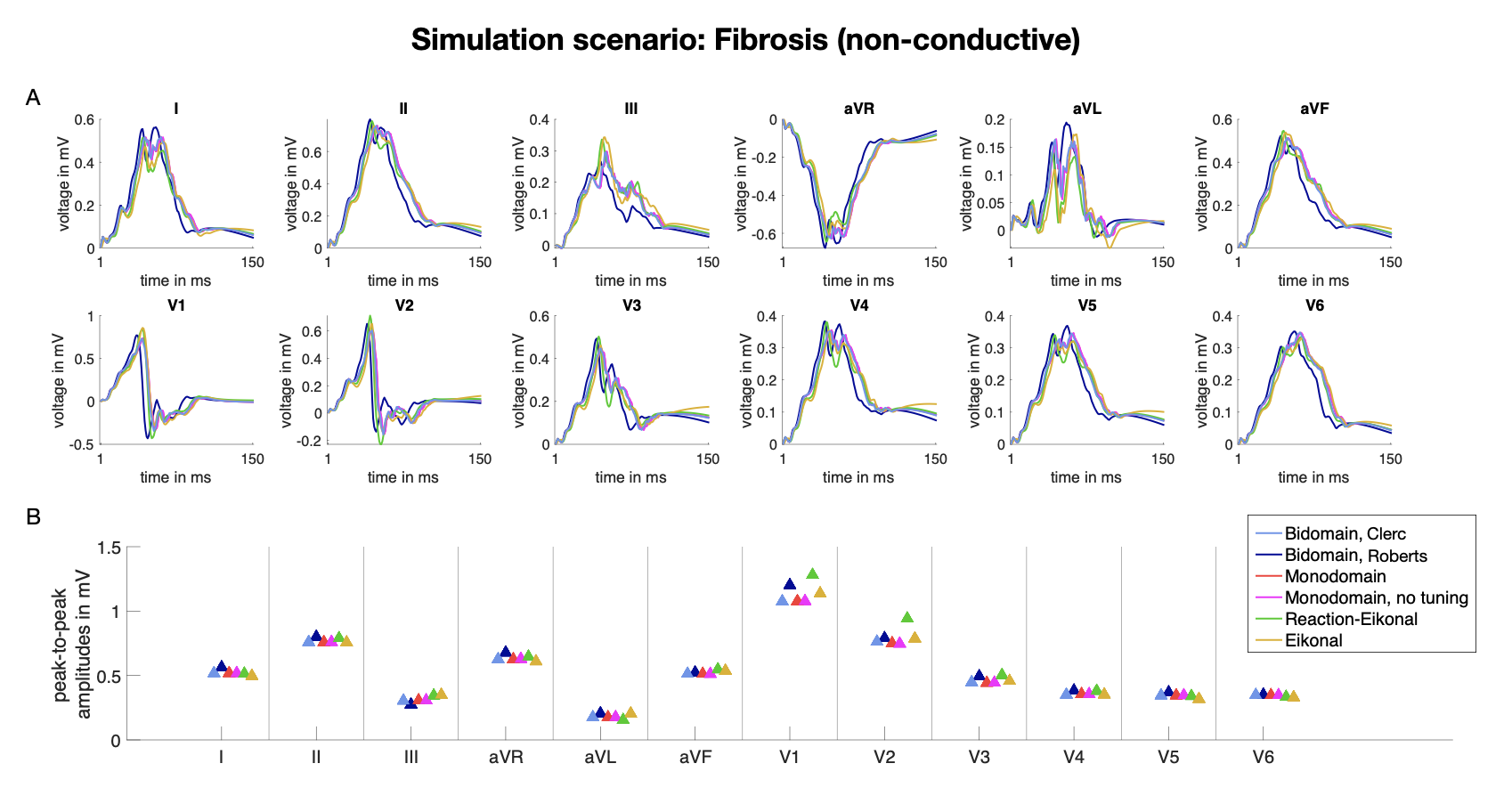}}
\caption{ECGs for different propagation models using the same forward calculation method (BEM) in the Interval [0, 150]\,ms (A) 
in the case of fibrosis modeled as slow conducting patches. Panel B shows the peak-to-peak amplitude features resulting for each propagation model. }
\label{fibextra_ecg_PD}
\end{figure*}

\begin{figure*}[!h]
\centerline{\includegraphics[width = \textwidth]{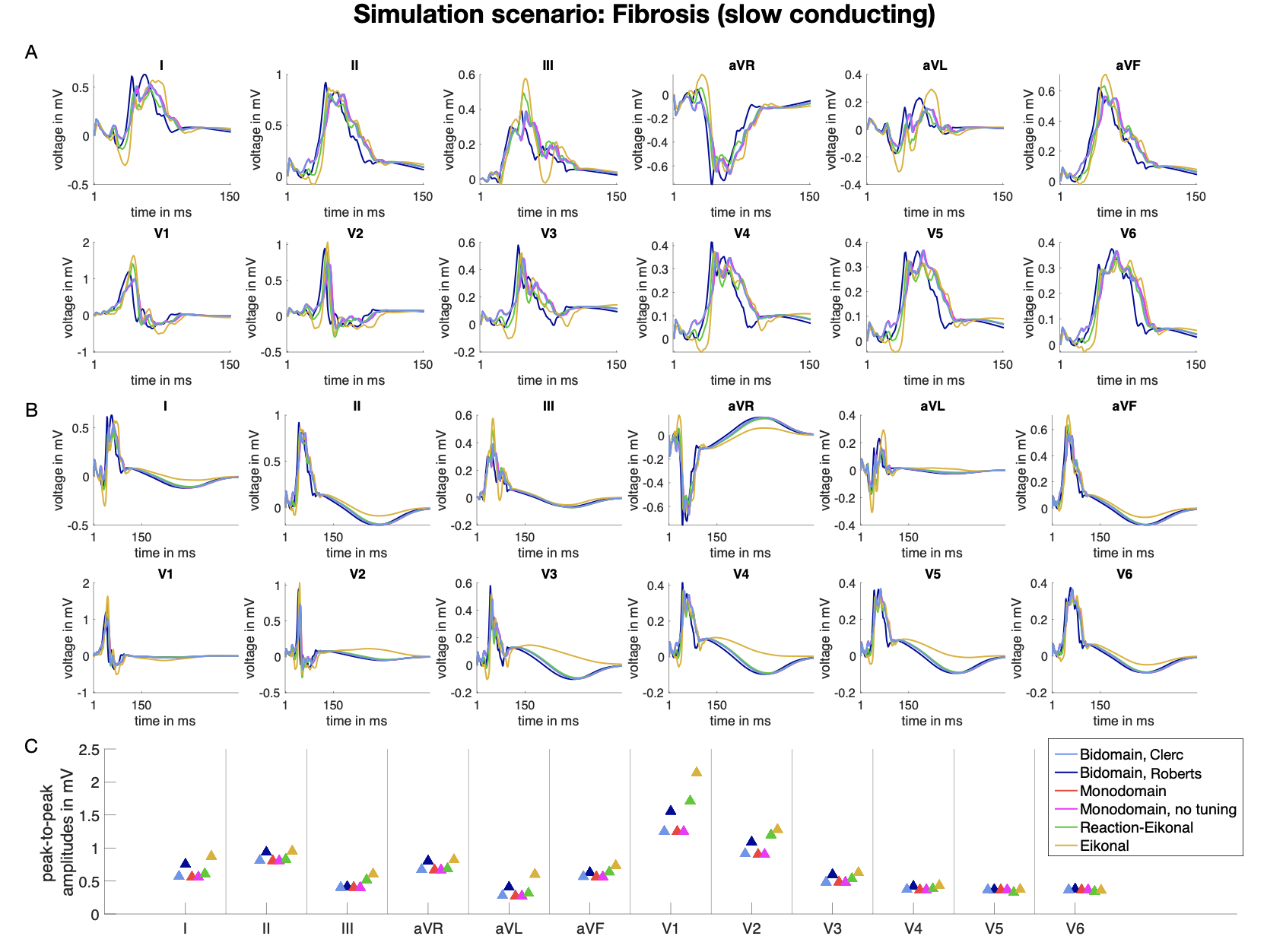}}
\caption{ECGs for different propagation models using the same forward calculation method (BEM) in the Interval [0, 150]\,ms (A) and [0, 650]\,ms (B) in the case of fibrosis modeled as slow conducting patches. Panel C shows the peak-to-peak amplitude features resulting for each propagation model. }
\label{fibslow_ecg_PD}
\end{figure*}

\begin{figure*}[!h]
\centerline{\includegraphics[width = \textwidth]{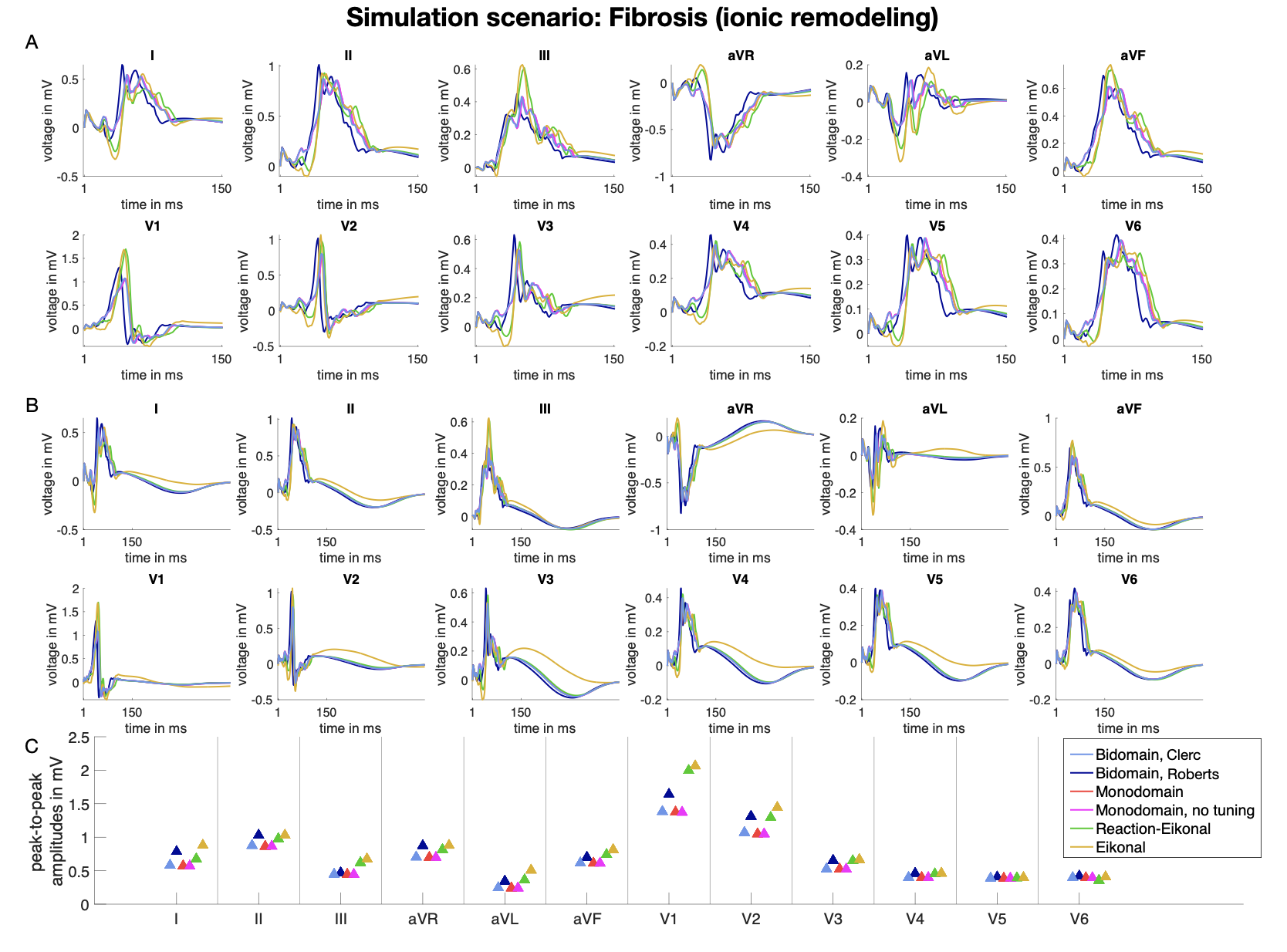}}
\caption{ECGs for different propagation models using the same forward calculation method (BEM) in the Interval [0, 150]\,ms (A) and [0, 650]\,ms (B) in the case of fibrosis modeled as ionic conductance rescaling. Panel C shows the peak-to-peak amplitude features resulting for each propagation model. }
\label{ecg_pd_fibremod}
\end{figure*}
\FloatBarrier
\clearpage
\section*{Effect of Forward Calculation Methods on ECGs and ECG Features in Different Simulation Scenarios}
\begin{figure*}[!h]
\centerline{\includegraphics[width = \textwidth]{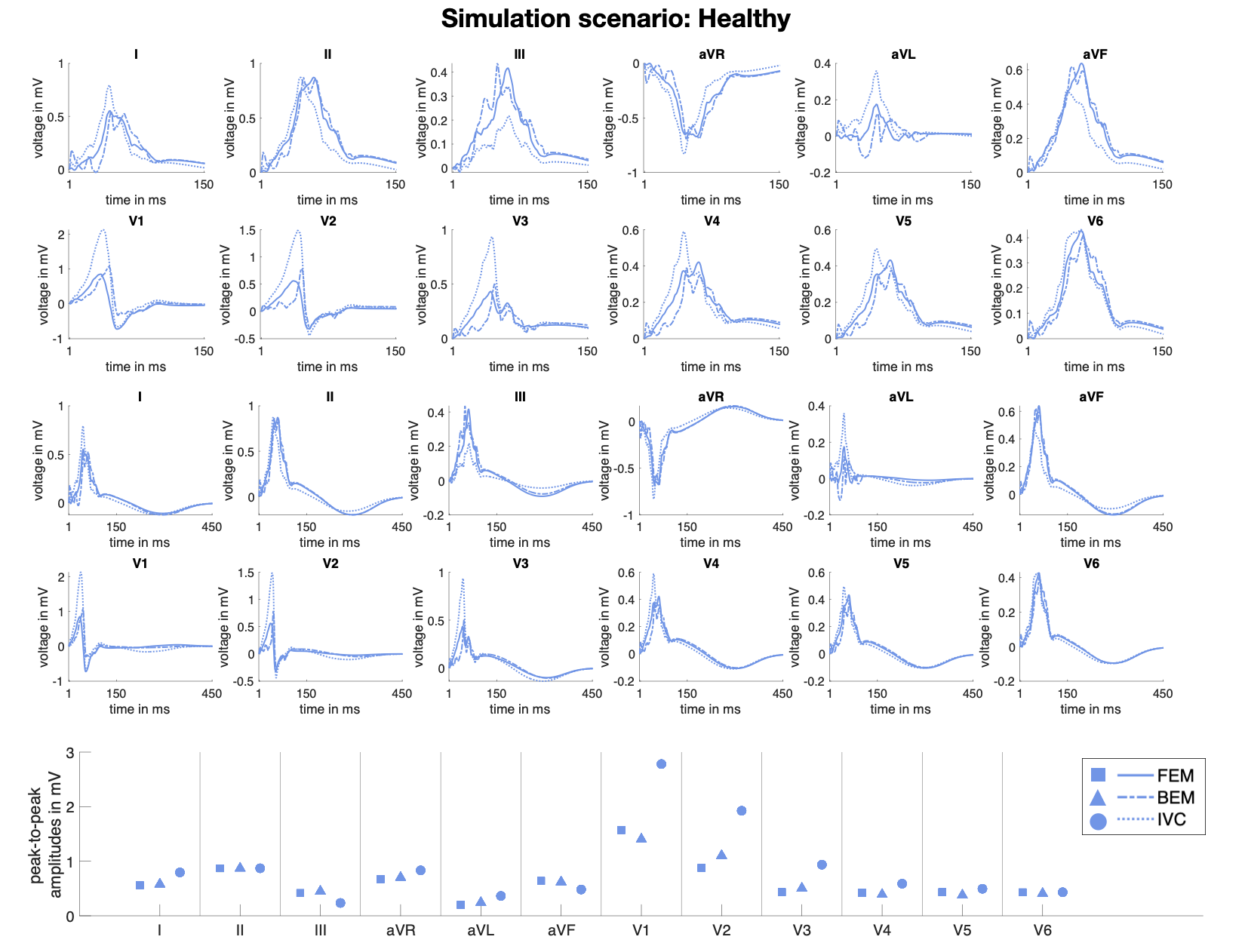}}
\caption{ECGs for different forward calculation methods using the same propagation model (bidomain with Clerc conductivity ratios) in the Interval [0, 150]\,ms (A) and [0, 650]\,ms (B) in the healthy case. Panel C shows the peak-to-peak amplitude features resulting for each forward calculation method. }
\label{ecg_fwc_healthy}
\end{figure*}

\begin{figure*}[!h]
\centerline{\includegraphics[width = \textwidth]{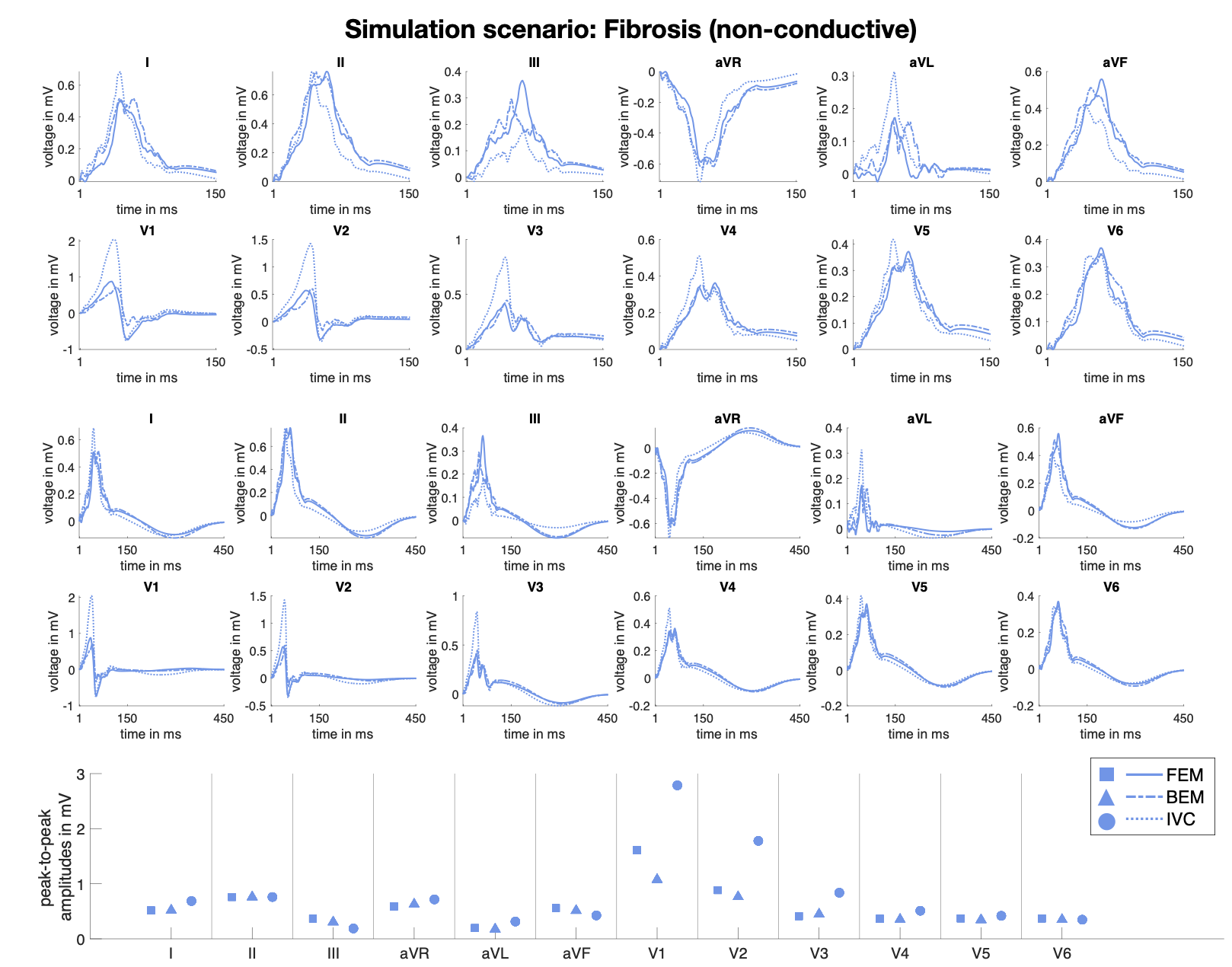}}
\caption{ECGs for different forward calculation methods using the same propagation model (bidomain with Clerc conductivity ratios) in the Interval [0, 150]\,ms (A) and [0, 650]\,ms (B) in the the case of fibrosis modeled as non-conductive tissue. Panel C shows the peak-to-peak amplitude features resulting for each forward calculation method. }
\label{ecg_pd_fibextra}
\end{figure*}

\begin{figure*}[!h]
\centerline{\includegraphics[width = \textwidth]{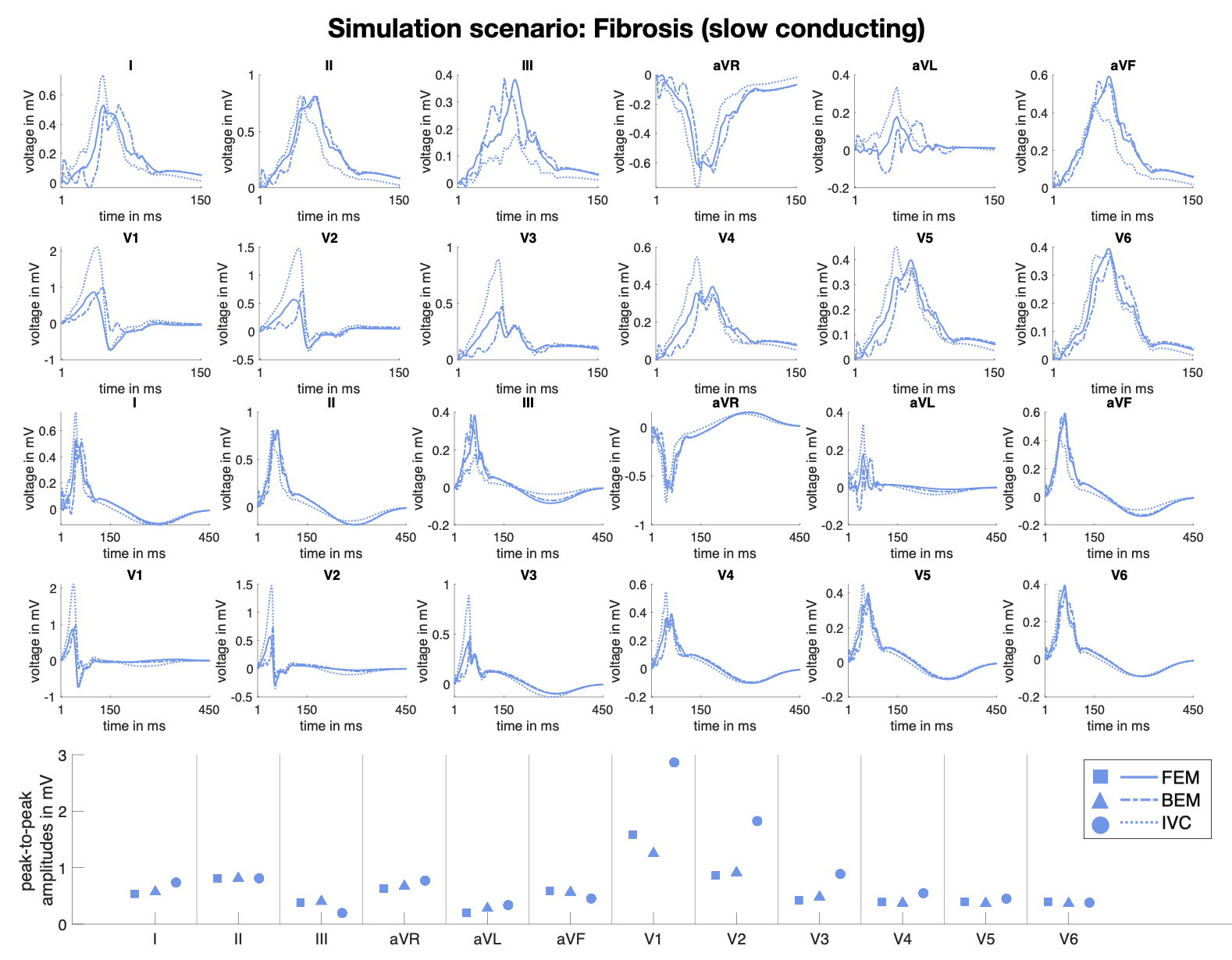}}
\caption{ECGs for different forward calculation methods using the same propagation model (bidomain with Clerc conductivity ratios) in the Interval [0, 150]\,ms (A) and [0, 650]\,ms (B) in the case of fibrosis modeled as slow conducting patches. Panel C shows the peak-to-peak amplitude features resulting for each forward calculation method. }
\label{ecg_pd_fibslow}
\end{figure*}

\begin{figure*}[!h]
\centerline{\includegraphics[width = \textwidth]{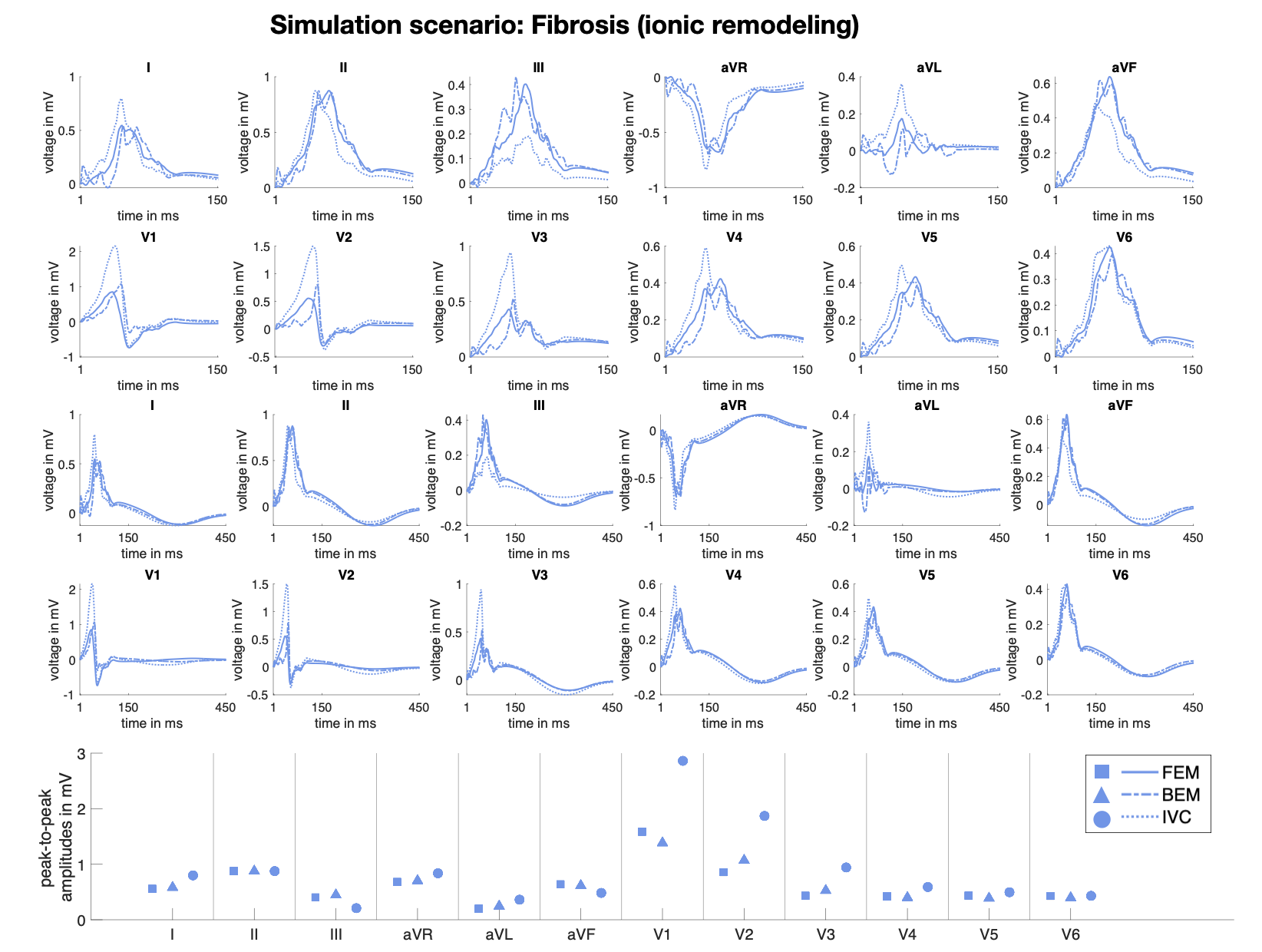}}
\caption{ECGs for different forward calculation methods using the same propagation model (bidomain with Clerc conductivity ratios) in the Interval [0, 150]\,ms (A) and [0, 650]\,ms (B) in the the case of fibrosis modeled as ionic conductance rescaling. Panel C shows the peak-to-peak amplitude features resulting for each forward calculation method. }
\label{ecg_fwc_fibremod}
\end{figure*}
\FloatBarrier
\clearpage
\section*{Effect of Fibrosis Remodeling Approach on ECGs}
\begin{figure*}[!h]
\centerline{\includegraphics[width = \textwidth]{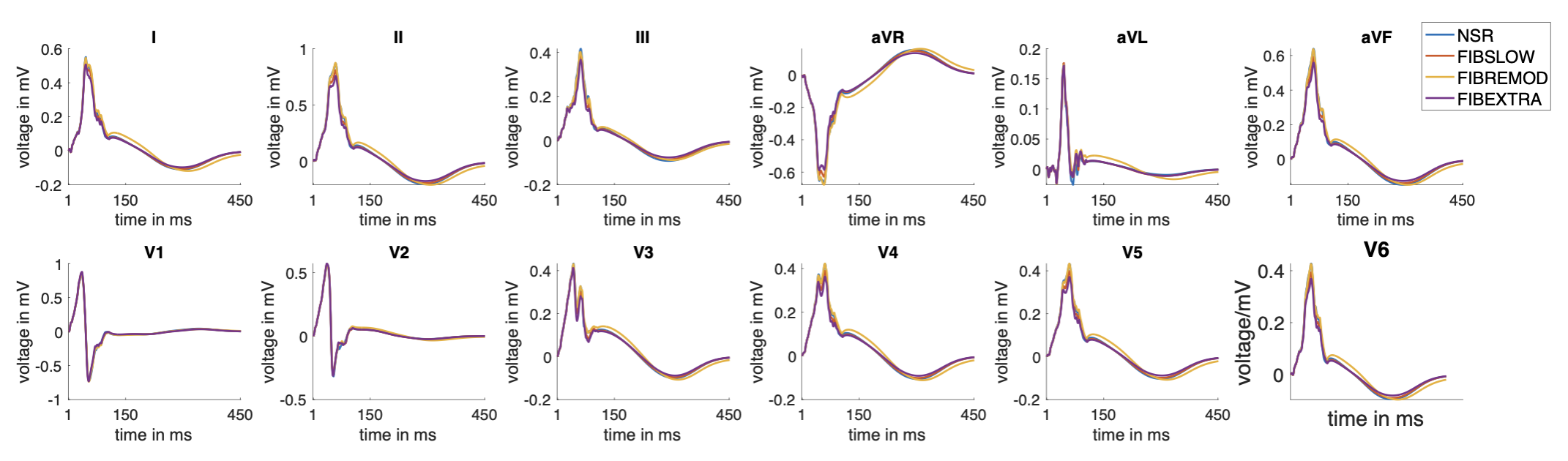}}
\caption{ECGs for different fibrosis modeling approaches. The blue, red, yellow and purple curve show the 12-lead surface ECG for the healthy case and the fibrotically infiltrated atrial geometry with fibrosis modeled as slow conducting tissue, ionic conductance rescaling and the percolation approach.}
\label{ecg_differentFibrosisModeling}
\end{figure*}

\end{document}